\documentclass[a4paper,11pt]{article}
\usepackage{jheppub} % for details on the use of the package, please
\usepackage[T1]{fontenc} % if needed
\usepackage{amsmath,amsthm}
\usepackage{amssymb,amsfonts}
\usepackage{graphicx}
\usepackage{dsfont}
\usepackage{relsize}
\usepackage{stmaryrd}
\usepackage{lmodern}
\usepackage{slantsc}
\usepackage{scalefnt}
\usepackage{subfigure}
\usepackage{xspace}
\usepackage{boxedminipage}
\usepackage{ifpdf}
\usepackage{multirow}
\usepackage{xifthen}
\usepackage{tensor}
\usepackage{array}
\usepackage{tabu}
\usepackage{tikz}
\usepackage{color}
\usepackage{diagbox}
\usetikzlibrary{arrows,matrix,calc,scopes,decorations.markings}
\allowdisplaybreaks[1]
\usepackage[mathcal]{euscript}

\newcommand{\be}{\begin{equation}}
\newcommand{\ee}{\end{equation}}
\newcommand{\bse}{\begin{subequations}}
\newcommand{\ese}{\end{subequations}}
\newcommand{\ket}[1]{\left|{#1}\right\rangle}
\newcommand{\bra}[1]{\left\langle#1\right|}
\newcommand{\braket}[2]{\left\langle #1 \middle| #2 \right\rangle}
\newcommand{\ketbra}[2]{\left| #1 \middle\rangle \middle\langle #2 \right|}

\newcommand{\bket}[1]{\Biggl|{#1}\Biggr\rangle}

\newcommand{\Z}{\mathbb{Z}}

\newcommand{\inv}[1]{{{#1}^{-1}}}
\newcommand{\ordG}{{|G|}}

\newcommand{\e}{\mathrm{e}}
\newcommand{\Hil}{\mathcal{H}}

\newcommand{\bpm}{\begin{pmatrix}}
\newcommand{\epm}{\end{pmatrix}}
\newcommand{\bmm}{\begin{matrix}}
\newcommand{\emm}{\end{matrix}}

\tikzset{snake it/.style={decorate, decoration={snake,amplitude=0.15mm,segment length=1mm}}}
\tikzset{->-/.style={decoration={
                        markings,
                        mark=at position .55 with {\arrow{latex}}},postaction={decorate}}}

\newtabulinestyle{dash=off 2pt}

 {\null\,\vcenter\bgroup\scriptsize
  \arraycolsep=.13885em
  \hbox\bgroup$\array{@{}#1@{}}}
 {\endarray$\egroup\egroup\,\null}

\newcolumntype{C}{>{\centering\arraybackslash} m{1.5em} }

\makeatletter
\newcommand*{\Relbarfill@}{\arrowfill@\Relbar\Relbar\Relbar}
\newcommand*{\xeq}[2][]{\ext@arrow 0055\Relbarfill@{#1}{#2}}
\makeatother

\usetikzlibrary{decorations.markings}
\usetikzlibrary{decorations.pathmorphing,positioning}
\tikzset{>=latex}
\tikzset{snake it/.style={decorate, decoration={snake,amplitude=0.15mm,segment length=1mm}}}
\tikzset{->-/.style={decoration={
			 markings,
			 mark=at position .55 with {\arrow{>}}},postaction={decorate}}}

\tikzset{-<-/.style={decoration={
			 markings,
			 mark=at position .55 with {\arrow{<}}},postaction={decorate}}}

\newcommand{\CylinderAA}{
\begin{tikzpicture}[baseline={([yshift=-.8ex]current bounding box.center)},scale=1]
\draw [] (2.1,8.2) ..controls (2.125,8.129) and (2.15,8.057) .. (2.3,8.)
			 ..controls (2.45,7.943) and (2.725,7.9) .. (3.,7.9)
			 ..controls (3.275,7.9) and (3.55,7.943) .. (3.7,8.)
			 ..controls (3.85,8.057) and (3.875,8.129) .. (3.9,8.2);
\draw [dashed] (2.1,8.2) ..controls (2.125,8.271) and (2.15,8.343) .. (2.3,8.4)
			 ..controls (2.45,8.457) and (2.725,8.5) .. (3.,8.5)
			 ..controls (3.275,8.5) and (3.55,8.457) .. (3.7,8.4)
			 ..controls (3.85,8.343) and (3.875,8.271) .. (3.9,8.2);
\draw [] (2.1,10.1) ..controls (2.1,10.029) and (2.143,9.957) .. (2.3,9.9)
			 ..controls (2.457,9.843) and (2.729,9.8) .. (3.,9.8)
			 ..controls (3.271,9.8) and (3.543,9.843) .. (3.7,9.9)
			 ..controls (3.857,9.957) and (3.9,10.029) .. (3.9,10.1)
			 ..controls (3.9,10.171) and (3.857,10.243) .. (3.7,10.3)
			 ..controls (3.543,10.357) and (3.271,10.4) .. (3.,10.4)
			 ..controls (2.729,10.4) and (2.457,10.357) .. (2.3,10.3)
			 ..controls (2.143,10.243) and (2.1,10.171) .. (2.1,10.1);
\draw [very thick] (2.3,9.9) ..controls (2.533,9.85) and (2.767,9.8) .. (3.,9.8)
			 ..controls (3.233,9.8) and (3.467,9.85) .. (3.7,9.9);
\draw [very thick] (2.3,8.) ..controls (2.533,7.95) and (2.767,7.9) .. (3.,7.9)
			 ..controls (3.233,7.9) and (3.467,7.95) .. (3.7,8.);
\draw [] (2.1,8.2) -- (2.1,10.1);
\draw [] (3.9,8.2) -- (3.9,9.322);
\draw [] (3.9,9.522) -- (3.9,10.1);
\draw [very thick] (2.3,9.9) -- (2.3,8.);
\draw [very thick] (3.7,9.9) -- (3.7,9.433);
\draw [very thick] (3.7,9.233) -- (3.7,8.);
\draw [->] (4.3,9.6) -- (3.4,9.2);
\draw [->] (1.7,9.3) -- (2.,9.);
\node [rotate=0.] at (4.5,9.6) {\scriptsize $A$};
\node [rotate=0.] at (1.6,9.4) {\scriptsize $B$};
\end{tikzpicture}
}

\newcommand{\CylinderAAState}{
\begin{tikzpicture}[baseline={([yshift=-.8ex]current bounding box.center)},scale=1]
\draw [->] (2.5,9.) ..controls (2.633,9.3) and (2.767,9.6) .. (2.9,9.8)
			 ..controls (3.033,10.) and (3.167,10.1) .. (3.3,10.2);
\draw [->] (2.5,9.) ..controls (2.683,9.158) and (2.867,9.317) .. (3.,9.4)
			 ..controls (3.133,9.483) and (3.217,9.492) .. (3.3,9.5);
\draw [->] (2.5,9.) ..controls (2.633,9.) and (2.767,9.) .. (2.9,9.)
			 ..controls (3.033,9.) and (3.167,9.) .. (3.3,9.);
\draw [->] (2.5,9.) ..controls (2.533,8.792) and (2.567,8.583) .. (2.7,8.4)
			 ..controls (2.833,8.217) and (3.067,8.058) .. (3.3,7.9);
\draw [->] (2.5,9.) ..controls (2.467,9.25) and (2.433,9.5) .. (2.3,9.7)
			 ..controls (2.167,9.9) and (1.933,10.05) .. (1.7,10.2);
\draw [->] (2.5,9.) ..controls (2.417,9.108) and (2.333,9.217) .. (2.2,9.3)
			 ..controls (2.067,9.383) and (1.883,9.442) .. (1.7,9.5);
\draw [->] (2.5,9.) ..controls (2.367,9.) and (2.233,9.) .. (2.1,9.)
			 ..controls (1.967,9.) and (1.833,9.) .. (1.7,9.);
\draw [->] (2.5,9.) ..controls (2.367,8.742) and (2.233,8.483) .. (2.1,8.3)
			 ..controls (1.967,8.117) and (1.833,8.008) .. (1.7,7.9);
\draw [->] (4.3,9.) ..controls (4.433,9.3) and (4.567,9.6) .. (4.7,9.8)
			 ..controls (4.833,10.) and (4.967,10.1) .. (5.1,10.2);
\draw [->] (4.3,9.) ..controls (4.483,9.158) and (4.667,9.317) .. (4.8,9.4)
			 ..controls (4.933,9.483) and (5.017,9.492) .. (5.1,9.5);
\draw [->] (4.3,9.) ..controls (4.433,9.) and (4.567,9.) .. (4.7,9.)
			 ..controls (4.833,9.) and (4.967,9.) .. (5.1,9.);
\draw [->] (4.3,9.) ..controls (4.333,8.792) and (4.367,8.583) .. (4.5,8.4)
			 ..controls (4.633,8.217) and (4.867,8.058) .. (5.1,7.9);
\draw [->] (4.3,9.) ..controls (4.267,9.25) and (4.233,9.5) .. (4.1,9.7)
			 ..controls (3.967,9.9) and (3.733,10.05) .. (3.5,10.2);
\draw [->] (4.3,9.) ..controls (4.217,9.108) and (4.133,9.217) .. (4.,9.3)
			 ..controls (3.867,9.383) and (3.683,9.442) .. (3.5,9.5);
\draw [->] (4.3,9.) ..controls (4.167,9.) and (4.033,9.) .. (3.9,9.)
			 ..controls (3.767,9.) and (3.633,9.) .. (3.5,9.);
\draw [->] (4.3,9.) ..controls (4.167,8.742) and (4.033,8.483) .. (3.9,8.3)
			 ..controls (3.767,8.117) and (3.633,8.008) .. (3.5,7.9);
\draw [dashed] (3.4,10.2) -- (3.4,7.9);
\draw [dashed] (1.6,10.2) -- (1.6,7.9);
\draw [dashed] (5.2,10.2) -- (5.2,7.9);
\draw [fill,rotate around={0.:(2.5,9.)}] (2.5,9.) ++(0.:0.08 and 0.08) arc(0.:360.:0.08 and 0.08);
\draw [fill,rotate around={0.:(4.3,9.)}] (4.3,9.) ++(0.:0.08 and 0.08) arc(0.:360.:0.08 and 0.08);
\node [rotate=0.] at (5.,10.4) {\scriptsize $k_1$};
\node [rotate=0.] at (1.9,10.4) {\scriptsize $k_1h$};
\node [rotate=0.] at (2.,7.8) {\scriptsize $k_2h$};
\node [rotate=0.] at (1.9,9.7) {\scriptsize ${g_{1}h}$};
\node [rotate=0.] at (1.9,9.2) {\scriptsize ${g_{2}h}$};
\node [rotate=0.] at (5.,9.7) {\scriptsize $g_{1}$};
\node [rotate=0.] at (5.,9.2) {\scriptsize $g_{2}$};
\node [rotate=0.] at (4.9,7.8) {\scriptsize $k_2$};
\node [rotate=0.] at (3.6,10.4) {\scriptsize $k_3$};
\node [rotate=0.] at (3.7,7.8) {\scriptsize $k_4$};
\node [rotate=0.] at (3.1,10.4) {\scriptsize $k_3h$};
\node [rotate=0.] at (3.1,7.8) {\scriptsize $k_4h$};
\node [rotate=0.] at (3.1,9.7) {\scriptsize $g_3h$};
\node [rotate=0.] at (3.1,9.2) {\scriptsize $g_4h$};
\node [rotate=0.] at (3.7,9.7) {\scriptsize $g_3$};
\node [rotate=0.] at (3.7,9.2) {\scriptsize $g_4$};
\end{tikzpicture}
}

\newcommand{\CylinderAB}{
\begin{tikzpicture}[baseline={([yshift=-.8ex]current bounding box.center)},scale=1]
\draw [] (2.2,7.8) ..controls (2.225,7.729) and (2.25,7.657) .. (2.4,7.6)
			 ..controls (2.55,7.543) and (2.825,7.5) .. (3.1,7.5)
			 ..controls (3.375,7.5) and (3.65,7.543) .. (3.8,7.6)
			 ..controls (3.95,7.657) and (3.975,7.729) .. (4.,7.8);
\draw [dashed] (2.2,7.8) ..controls (2.225,7.871) and (2.25,7.943) .. (2.4,8.)
			 ..controls (2.55,8.057) and (2.825,8.1) .. (3.1,8.1)
			 ..controls (3.375,8.1) and (3.65,8.057) .. (3.8,8.)
			 ..controls (3.95,7.943) and (3.975,7.871) .. (4.,7.8);
\draw [] (2.2,10.1) ..controls (2.2,10.029) and (2.243,9.957) .. (2.4,9.9)
			 ..controls (2.557,9.843) and (2.829,9.8) .. (3.1,9.8)
			 ..controls (3.371,9.8) and (3.643,9.843) .. (3.8,9.9)
			 ..controls (3.957,9.957) and (4.,10.029) .. (4.,10.1)
			 ..controls (4.,10.171) and (3.957,10.243) .. (3.8,10.3)
			 ..controls (3.643,10.357) and (3.371,10.4) .. (3.1,10.4)
			 ..controls (2.829,10.4) and (2.557,10.357) .. (2.4,10.3)
			 ..controls (2.243,10.243) and (2.2,10.171) .. (2.2,10.1);
\draw [very thick] (2.2,9.4) ..controls (2.225,9.329) and (2.25,9.257) .. (2.4,9.2)
			 ..controls (2.55,9.143) and (2.825,9.1) .. (3.1,9.1)
			 ..controls (3.375,9.1) and (3.65,9.143) .. (3.8,9.2)
			 ..controls (3.95,9.257) and (3.975,9.329) .. (4.,9.4);
\draw [very thick] (2.2,8.5) ..controls (2.225,8.429) and (2.25,8.357) .. (2.4,8.3)
			 ..controls (2.55,8.243) and (2.825,8.2) .. (3.1,8.2)
			 ..controls (3.375,8.2) and (3.65,8.243) .. (3.8,8.3)
			 ..controls (3.95,8.357) and (3.975,8.429) .. (4.,8.5);
\draw [very thick, dashed] (2.2,9.4) ..controls (2.225,9.471) and (2.25,9.543) .. (2.4,9.6)
			 ..controls (2.55,9.657) and (2.825,9.7) .. (3.1,9.7)
			 ..controls (3.375,9.7) and (3.65,9.657) .. (3.8,9.6)
			 ..controls (3.95,9.543) and (3.975,9.471) .. (4.,9.4);
\draw [very thick, dashed] (2.2,8.5) ..controls (2.225,8.571) and (2.25,8.643) .. (2.4,8.7)
			 ..controls (2.55,8.757) and (2.825,8.8) .. (3.1,8.8)
			 ..controls (3.375,8.8) and (3.65,8.757) .. (3.8,8.7)
			 ..controls (3.95,8.643) and (3.975,8.571) .. (4.,8.5);
\draw [] (2.2,7.8) -- (2.2,10.1);
\draw [] (4.,7.8) -- (4.,10.1);
\draw [->] (4.6,9.2) -- (4.1,9.);
\draw [->] (1.7,9.2) -- (2.1,9.8);
\draw [->] (1.7,8.8) -- (2.,8.3);
\draw [fill,rotate around={0.:(3.8,9.9)}] (3.8,9.9) ++(0.:0.08 and 0.08) arc(0.:360.:0.08 and 0.08);
\draw [fill,rotate around={0.:(3.8,7.6)}] (3.8,7.6) ++(0.:0.08 and 0.08) arc(0.:360.:0.08 and 0.08);
\draw [fill,rotate around={0.:(3.8,9.)}] (3.8,9.) ++(0.:0.08 and 0.08) arc(0.:360.:0.08 and 0.08);
\node [rotate=0.] at (4.8,9.3) {\scriptsize $A$};
\node [rotate=0.] at (1.6,9.) {\scriptsize $B$};
\node [rotate=0.] at (3.5,10.) {\scriptsize $v_{B1}$};
\node [rotate=0.] at (3.5,7.7) {\scriptsize $v_{B2}$};
\node [rotate=0.] at (3.5,9.) {\scriptsize $v_A$};
\end{tikzpicture}
}

\newcommand{\CylinderABState}{
\begin{tikzpicture}[baseline={([yshift=-.8ex]current bounding box.center)},scale=1]
\draw [->] (3.2,8.7) ..controls (2.933,8.775) and (2.667,8.85) .. (2.5,9.)
			 ..controls (2.333,9.15) and (2.267,9.375) .. (2.2,9.6);
\draw [->] (3.2,8.7) ..controls (3.042,8.875) and (2.883,9.05) .. (2.8,9.2)
			 ..controls (2.717,9.35) and (2.708,9.475) .. (2.7,9.6);
\draw [->] (3.2,8.7) ..controls (3.492,8.675) and (3.783,8.65) .. (4.,8.8)
			 ..controls (4.217,8.95) and (4.358,9.275) .. (4.5,9.6);
\draw [->] (3.2,8.7) ..controls (2.883,8.7) and (2.567,8.7) .. (2.4,8.6)
			 ..controls (2.233,8.5) and (2.217,8.3) .. (2.2,8.1);
\draw [->] (3.2,8.7) ..controls (3.033,8.6) and (2.867,8.5) .. (2.8,8.4)
			 ..controls (2.733,8.3) and (2.767,8.2) .. (2.8,8.1);
\draw [->] (3.2,8.7) ..controls (3.442,8.65) and (3.683,8.6) .. (3.9,8.5)
			 ..controls (4.117,8.4) and (4.308,8.25) .. (4.5,8.1);
\draw [->] (3.2,7.3) ..controls (2.933,7.3) and (2.667,7.3) .. (2.5,7.4)
			 ..controls (2.333,7.5) and (2.267,7.7) .. (2.2,7.9);
\draw [->] (3.2,7.3) ..controls (3.083,7.35) and (2.967,7.4) .. (2.9,7.5)
			 ..controls (2.833,7.6) and (2.817,7.75) .. (2.8,7.9);
\draw [->] (3.2,7.3) ..controls (3.542,7.3) and (3.883,7.3) .. (4.1,7.4)
			 ..controls (4.317,7.5) and (4.408,7.7) .. (4.5,7.9);
\draw [->] (3.1,10.4) ..controls (2.875,10.35) and (2.65,10.3) .. (2.5,10.2)
			 ..controls (2.35,10.1) and (2.275,9.95) .. (2.2,9.8);
\draw [->] (3.1,10.4) ..controls (2.983,10.3) and (2.867,10.2) .. (2.8,10.1)
			 ..controls (2.733,10.) and (2.717,9.9) .. (2.7,9.8);
\draw [->] (3.1,10.4) ..controls (3.433,10.35) and (3.767,10.3) .. (4.,10.2)
			 ..controls (4.233,10.1) and (4.367,9.95) .. (4.5,9.8);
\draw [dashed] (1.6,9.7) -- (4.8,9.7);
\draw [dashed] (1.6,8.) -- (4.8,8.);
\draw [fill,rotate around={0.:(3.1,10.4)}] (3.1,10.4) ++(0.:0.08 and 0.08) arc(0.:360.:0.08 and 0.08);
\draw [fill,rotate around={0.:(3.2,8.7)}] (3.2,8.7) ++(0.:0.08 and 0.08) arc(0.:360.:0.08 and 0.08);
\draw [fill,rotate around={0.:(3.2,7.3)}] (3.2,7.3) ++(0.:0.08 and 0.08) arc(0.:360.:0.08 and 0.08);
\node [rotate=0.] at (2.,9.4) {\scriptsize $g_1h$};
\node [rotate=0.] at (3.,9.4) {\scriptsize $g_2h$};
\node [rotate=0.] at (1.9,8.3) {\scriptsize $g'_1h$};
\node [rotate=0.] at (3.1,8.3) {\scriptsize $g'_2h$};
\node [rotate=0.] at (1.9,7.7) {\scriptsize $g'_1k_2$};
\node [rotate=0.] at (3.1,7.7) {\scriptsize $g'_2k_2$};
\node [rotate=0.] at (2.,10.) {\scriptsize $g_1k_1$};
\node [rotate=0.] at (3.1,10.) {\scriptsize $g_2k_1$};
\end{tikzpicture}
}

\newcommand{\Decomposej}{
\begin{tikzpicture}[baseline={([yshift=-.8ex]current bounding box.center)},scale=1]
\draw [->-] (1.6,7.) -- (1.6,8.6);
\draw [->-] (1.6,8.6) -- (1.6,10.2);
\draw [->-] (4.8,6.8) -- (4.8,7.6);
\draw [->-] (4.8,7.6) -- (4.8,8.6);
\draw [->-] (4.8,8.6) -- (4.8,9.6);
\draw [->-] (4.8,9.6) -- (4.8,10.4);
\node [fill=white,draw,rounded corners=.06cm, circle, rotate=0.] at (1.6,8.6) {\scriptsize $k$};
\node [fill=white,draw,rounded corners=.06cm, circle, rotate=0.] at (4.8,8.6) {\scriptsize $k$};
\node [fill=white,draw,rounded corners=.06cm, rotate=0.] at (4.8,7.6) {\scriptsize $U^j_{p\alpha}$};
\node [fill=white,draw,rounded corners=.06cm, rotate=0.] at (4.8,9.6) {\scriptsize ${U^j_{p\alpha}}^\dagger$};
\node [rotate=0.] at (1.8,7.4) {\scriptsize $j$};
\node [rotate=0.] at (5.,7.2) {\scriptsize $j$};
\node [rotate=0.] at (5.,8.2) {\scriptsize $p$};
\node [rotate=0.] at (5.,9.) {\scriptsize $p$};
\node [rotate=0.] at (5.,10.2) {\scriptsize $j$};
\node [rotate=0.] at (1.8,9.8) {\scriptsize $j$};
\node [rotate=0.] at (3.2,8.6) {\scriptsize $=\sum_p\sum_{\alpha\in{N^j_p}}$};
\end{tikzpicture}
}

\newcommand{\Disk}{
\begin{tikzpicture}[baseline={([yshift=-.8ex]current bounding box.center)},scale=1,baseline=210]
\draw [lightgray,fill,rotate around={0.:(2.6,9.2)}] (2.6,9.2) ++(0.:2. and 2.) arc(0.:360.:2. and 2.);
\draw [very thick] (2.6,10.2) ..controls (2.857,10.2) and (3.114,10.086) .. (3.3,9.9)
			 ..controls (3.486,9.714) and (3.6,9.457) .. (3.6,9.2)
			 ..controls (3.6,8.943) and (3.486,8.686) .. (3.3,8.5)
			 ..controls (3.114,8.314) and (2.857,8.2) .. (2.6,8.2)
			 ..controls (2.343,8.2) and (2.086,8.314) .. (1.9,8.5)
			 ..controls (1.714,8.686) and (1.6,8.943) .. (1.6,9.2)
			 ..controls (1.6,9.457) and (1.714,9.714) .. (1.9,9.9)
			 ..controls (2.086,10.086) and (2.343,10.2) .. (2.6,10.2);
\draw [rotate around={0.:(2.6,9.2)}] (2.6,9.2) ++(0.:2. and 2.) arc(0.:360.:2. and 2.);
\draw [rotate around={0.:(2.6,9.2)}] (2.6,9.2) ++(0.:1. and 1.) arc(0.:360.:1. and 1.);
\draw [fill,rotate around={0.:(2.6,9.2)}] (2.6,9.2) ++(0.:0.08 and 0.08) arc(0.:360.:0.08 and 0.08);
\node [rotate=0.] at (2.9,9.7) {\scriptsize $A$};
\node [rotate=0.] at (3.5,10.4) {\scriptsize $B$};
\node [rotate=0.] at (4.7,9.6) {\scriptsize $K$};
\end{tikzpicture}
}

\newcommand{\FourValentState}[4]{
\begin{tikzpicture}[baseline={([yshift=-.8ex]current bounding box.center)},scale=1]
\draw [->-] (1.6,9.6) -- (2.4,9.6);
\draw [->-] (2.4,8.8) -- (2.4,9.6);
\draw [->-] (3.2,9.6) -- (2.4,9.6);
\draw [->-] (2.4,10.4) -- (2.4,9.6);
\node [rotate=0.] at (1.9,9.4) {\scriptsize $#1$};
\node [rotate=0.] at (2.6,9.1) {\scriptsize $#2$};
\node [rotate=0.] at (2.9,9.8) {\scriptsize $#3$};
\node [rotate=0.] at (2.2,10.1) {\scriptsize $#4$};
\end{tikzpicture}
}

\newcommand{\FusionBasis}{
\begin{tikzpicture}[baseline={([yshift=-.8ex]current bounding box.center)},scale=1]
\draw [->] (1.6,9.3) -- (1.6,10.1);
\draw [->] (2.4,9.3) -- (2.4,10.1);
\draw [->] (4.4,9.3) -- (4.4,10.1);
\draw [] (1.6,9.3) -- (2.4,9.3);
\draw [] (2.4,9.3) -- (3.,9.3);
\draw [] (3.8,9.3) -- (4.4,9.3);
\draw [dashed] (3.,9.3) -- (3.8,9.3);
\node [rotate=0.] at (1.6,10.4) {\scriptsize $j_1m_1$};
\node [rotate=0.] at (2.4,10.4) {\scriptsize $j_2m_2$};
\node [rotate=0.] at (4.4,10.4) {\scriptsize $j_Lm_L$};
\node [rotate=0.] at (3.,9.1) {\scriptsize $\eta$};
\end{tikzpicture}
}

\newcommand{\FusionBasisCylinder}{
\begin{tikzpicture}[baseline={([yshift=-.8ex]current bounding box.center)},scale=1]
\draw [->] (3.6,8.8) -- (3.6,9.6);
\draw [->] (4.4,8.8) -- (4.4,9.6);
\draw [->] (6.4,8.8) -- (6.4,9.6);
\draw [] (3.6,8.8) -- (4.4,8.8);
\draw [] (4.4,8.8) -- (5.,8.8);
\draw [] (5.8,8.8) -- (6.4,8.8);
\draw [dashed] (5.,8.8) -- (5.8,8.8);
\draw [] (6.4,8.8) -- (7.6,8.8);
\draw [->] (7.6,8.8) -- (7.6,9.6);
\draw [->] (8.4,8.8) -- (8.4,9.6);
\draw [->] (1.6,8.8) -- (1.6,9.6);
\draw [->] (2.4,8.8) -- (2.4,9.6);
\draw [] (7.6,8.8) -- (8.4,8.8);
\draw [] (1.6,8.8) -- (2.4,8.8);
\draw [] (2.4,8.8) -- (3.6,8.8);
\draw [fill,rotate around={0.:(7.,8.8)}] (7.,8.8) ++(0.:0.08 and 0.08) arc(0.:360.:0.08 and 0.08);
\draw [fill,rotate around={0.:(3.,8.8)}] (3.,8.8) ++(0.:0.08 and 0.08) arc(0.:360.:0.08 and 0.08);
\node [rotate=0.] at (3.6,9.9) {\scriptsize $j_1m_1$};
\node [rotate=0.] at (4.4,9.9) {\scriptsize $j_2m_2$};
\node [rotate=0.] at (6.4,9.9) {\scriptsize $j_Lm_L$};
\node [rotate=0.] at (5.,8.6) {\scriptsize $\xi$};
\node [rotate=0.] at (7.6,9.9) {\scriptsize $p_1x_1$};
\node [rotate=0.] at (8.4,9.9) {\scriptsize $p_3x_3$};
\node [rotate=0.] at (2.4,9.9) {\scriptsize $p_4x_4$};
\node [rotate=0.] at (1.6,9.9) {\scriptsize $p_2x_2$};
\end{tikzpicture}
}

\newcommand{\FusionBasisCylinderII}{
\begin{tikzpicture}[baseline={([yshift=-.8ex]current bounding box.center)},scale=1]
\draw [->] (3.2,9.2) -- (3.2,10.);
\draw [->] (6.4,9.2) -- (6.4,10.);
\draw [] (6.4,9.2) -- (8.,9.2);
\draw [->] (8.,9.2) -- (8.,10.);
\draw [->] (8.8,9.2) -- (8.8,10.);
\draw [->] (0.8,9.2) -- (0.8,10.);
\draw [->] (1.6,9.2) -- (1.6,10.);
\draw [] (8.,9.2) -- (8.8,9.2);
\draw [] (0.8,9.2) -- (1.6,9.2);
\draw [] (1.6,9.2) -- (3.2,9.2);
\draw [->-] (4.4,9.2) -- (5.2,9.2);
\draw [dashed] (3.6,9.2) -- (4.4,9.2);
\draw [dashed] (5.2,9.2) -- (6.,9.2);
\draw [] (3.2,9.2) -- (3.6,9.2);
\draw [] (6.,9.2) -- (6.4,9.2);
\node [rotate=0.] at (4.8,8.8) {\scriptsize $\eta_0$};
\node [rotate=0.] at (1.,10.4) {\scriptsize $j_1m_1$};
\node [rotate=0.] at (1.8,10.4) {\scriptsize $j_2m_2$};
\node [rotate=0.] at (3.4,10.4) {\scriptsize $j_{L_l}m_{L_l}$};
\node [rotate=0.] at (6.6,10.4) {\scriptsize $j_{L_l+1}m_{L_l+1}$};
\node [rotate=0.] at (9.,10.4) {\scriptsize $j_{L_l+L_r}m_{L_l+1+L_r}$};
\end{tikzpicture}
}

\newcommand{\GenericSurface}{
\begin{tikzpicture}[baseline={([yshift=-.8ex]current bounding box.center)},scale=1]
\draw [lightgray, fill] (2.5,10.4) ..controls (1.71,10.193) and (1.654,9.7) .. (1.6,9.3)
			 ..controls (1.546,8.9) and (1.493,8.593) .. (1.6,8.2)
			 ..controls (1.707,7.807) and (1.972,7.329) .. (2.4,7.1)
			 ..controls (2.828,6.871) and (3.419,6.892) .. (4.1,6.9)
			 ..controls (4.781,6.908) and (5.552,6.903) .. (6.2,7.)
			 ..controls (6.848,7.097) and (7.372,7.297) .. (7.6,7.8)
			 ..controls (7.828,8.303) and (7.76,9.108) .. (7.5,9.6)
			 ..controls (7.24,10.092) and (6.788,10.271) .. (5.8,10.4)
			 ..controls (4.812,10.529) and (3.29,10.607) .. (2.5,10.4);
\draw [white, fill] (2.9,10.) ..controls (2.675,9.95) and (2.8,9.65) .. (2.9,9.5)
			 ..controls (3.,9.35) and (3.075,9.35) .. (3.3,9.4)
			 ..controls (3.525,9.45) and (3.9,9.55) .. (3.8,9.7)
			 ..controls (3.7,9.85) and (3.125,10.05) .. (2.9,10.);
\draw [white, fill] (3.1,7.9) ..controls (2.9,7.775) and (2.8,7.45) .. (3.,7.3)
			 ..controls (3.2,7.15) and (3.7,7.175) .. (3.9,7.3)
			 ..controls (4.1,7.425) and (4.,7.65) .. (3.8,7.8)
			 ..controls (3.6,7.95) and (3.3,8.025) .. (3.1,7.9);
\draw [white, fill] (4.1,8.7) ..controls (4.1,8.575) and (4.325,8.325) .. (4.5,8.3)
			 ..controls (4.675,8.275) and (4.8,8.475) .. (4.8,8.6)
			 ..controls (4.8,8.725) and (4.675,8.775) .. (4.5,8.8)
			 ..controls (4.325,8.825) and (4.1,8.825) .. (4.1,8.7);
\draw [white, fill] (5.7,9.3) ..controls (5.6,9.075) and (5.7,8.75) .. (6.,8.7)
			 ..controls (6.3,8.65) and (6.8,8.875) .. (6.9,9.1)
			 ..controls (7.,9.325) and (6.7,9.55) .. (6.4,9.6)
			 ..controls (6.1,9.65) and (5.8,9.525) .. (5.7,9.3);
\draw [] (2.5,10.4) ..controls (1.71,10.193) and (1.654,9.7) .. (1.6,9.3)
			 ..controls (1.546,8.9) and (1.493,8.593) .. (1.6,8.2)
			 ..controls (1.707,7.807) and (1.972,7.329) .. (2.4,7.1)
			 ..controls (2.828,6.871) and (3.419,6.892) .. (4.1,6.9)
			 ..controls (4.781,6.908) and (5.552,6.903) .. (6.2,7.)
			 ..controls (6.848,7.097) and (7.372,7.297) .. (7.6,7.8)
			 ..controls (7.828,8.303) and (7.76,9.108) .. (7.5,9.6)
			 ..controls (7.24,10.092) and (6.788,10.271) .. (5.8,10.4)
			 ..controls (4.812,10.529) and (3.29,10.607) .. (2.5,10.4);
\draw [] (2.9,10.) ..controls (2.675,9.95) and (2.8,9.65) .. (2.9,9.5)
			 ..controls (3.,9.35) and (3.075,9.35) .. (3.3,9.4)
			 ..controls (3.525,9.45) and (3.9,9.55) .. (3.8,9.7)
			 ..controls (3.7,9.85) and (3.125,10.05) .. (2.9,10.);
\draw [] (3.1,7.9) ..controls (2.9,7.775) and (2.8,7.45) .. (3.,7.3)
			 ..controls (3.2,7.15) and (3.7,7.175) .. (3.9,7.3)
			 ..controls (4.1,7.425) and (4.,7.65) .. (3.8,7.8)
			 ..controls (3.6,7.95) and (3.3,8.025) .. (3.1,7.9);
\draw [] (5.7,9.3) ..controls (5.6,9.075) and (5.7,8.75) .. (6.,8.7)
			 ..controls (6.3,8.65) and (6.8,8.875) .. (6.9,9.1)
			 ..controls (7.,9.325) and (6.7,9.55) .. (6.4,9.6)
			 ..controls (6.1,9.65) and (5.8,9.525) .. (5.7,9.3);
\draw [] (4.5,8.8) ..controls (4.325,8.825) and (4.1,8.825) .. (4.1,8.7)
			 ..controls (4.1,8.575) and (4.325,8.325) .. (4.5,8.3)
			 ..controls (4.675,8.275) and (4.8,8.475) .. (4.8,8.6)
			 ..controls (4.8,8.725) and (4.675,8.775) .. (4.5,8.8);
\draw [very thick] (2.9,9.5) ..controls (2.833,9.283) and (2.767,9.067) .. (2.8,8.8)
			 ..controls (2.833,8.533) and (2.967,8.217) .. (3.1,7.9);
\draw [very thick] (3.9,7.3) ..controls (4.475,7.283) and (5.05,7.267) .. (5.4,7.5)
			 ..controls (5.75,7.733) and (5.875,8.217) .. (6.,8.7);
\draw [very thick] (3.8,9.7) ..controls (4.242,9.683) and (4.683,9.667) .. (5.,9.6)
			 ..controls (5.317,9.533) and (5.508,9.417) .. (5.7,9.3);
\draw [very thick] (4.1,9.2) ..controls (3.836,9.1) and (3.727,8.918) .. (3.7,8.7)
			 ..controls (3.673,8.482) and (3.727,8.227) .. (4.,8.1)
			 ..controls (4.273,7.973) and (4.764,7.973) .. (5.,8.2)
			 ..controls (5.236,8.427) and (5.218,8.882) .. (5.,9.1)
			 ..controls (4.782,9.318) and (4.364,9.3) .. (4.1,9.2);
\node [rotate=0.] at (3.3,9.) {\scriptsize $A$};
\node [rotate=0.] at (2.1,9.) {\scriptsize $B$};
\node [rotate=0.] at (4.6,9.) {\scriptsize $B$};
\node [rotate=0.] at (3.2,9.8) {\scriptsize $K_1$};
\node [rotate=0.] at (3.4,7.8) {\scriptsize $K_2$};
\node [rotate=0.] at (6.1,9.4) {\scriptsize $K_3$};
\node [rotate=0.] at (7.6,9.9) {\scriptsize $K_4$};
\node [rotate=0.] at (4.4,8.6) {\scriptsize $K_5$};
\end{tikzpicture}
}

\newcommand{\GenericSurfaceAA}{
\begin{tikzpicture}[baseline={([yshift=-.8ex]current bounding box.center)},scale=1]
\draw [very thick] (2.9,9.5) ..controls (2.833,9.283) and (2.767,9.067) .. (2.8,8.8)
			 ..controls (2.833,8.533) and (2.967,8.217) .. (3.1,7.9);
\draw [very thick] (3.9,7.3) ..controls (4.475,7.283) and (5.05,7.267) .. (5.4,7.5)
			 ..controls (5.75,7.733) and (5.875,8.217) .. (6.,8.7);
\draw [very thick] (3.8,9.7) ..controls (4.242,9.683) and (4.683,9.667) .. (5.,9.6)
			 ..controls (5.317,9.533) and (5.508,9.417) .. (5.7,9.3);
\draw [very thick] (4.1,9.2) ..controls (3.836,9.1) and (3.727,8.918) .. (3.7,8.7)
			 ..controls (3.673,8.482) and (3.727,8.227) .. (4.,8.1)
			 ..controls (4.273,7.973) and (4.764,7.973) .. (5.,8.2)
			 ..controls (5.236,8.427) and (5.218,8.882) .. (5.,9.1)
			 ..controls (4.782,9.318) and (4.364,9.3) .. (4.1,9.2);
\draw [] (2.9,9.5) ..controls (2.916,9.46) and (2.932,9.42) .. (3.,9.4)
			 ..controls (3.068,9.38) and (3.189,9.379) .. (3.3,9.4)
			 ..controls (3.411,9.421) and (3.511,9.463) .. (3.6,9.5)
			 ..controls (3.689,9.537) and (3.768,9.568) .. (3.8,9.6)
			 ..controls (3.832,9.632) and (3.816,9.666) .. (3.8,9.7);
\draw [] (5.7,9.3) ..controls (5.693,9.193) and (5.687,9.087) .. (5.7,9.)
			 ..controls (5.713,8.913) and (5.747,8.847) .. (5.8,8.8)
			 ..controls (5.853,8.753) and (5.927,8.727) .. (6.,8.7);
\draw [] (3.1,7.9) ..controls (3.187,7.96) and (3.275,8.021) .. (3.4,8.)
			 ..controls (3.525,7.979) and (3.688,7.878) .. (3.8,7.8)
			 ..controls (3.912,7.722) and (3.972,7.667) .. (4.,7.6)
			 ..controls (4.028,7.533) and (4.022,7.452) .. (4.,7.4)
			 ..controls (3.978,7.348) and (3.939,7.324) .. (3.9,7.3);
\node [rotate=0.] at (3.3,9.) {\scriptsize $A$};
\node [rotate=0.] at (3.,9.8) {\scriptsize $K_1$};
\node [rotate=0.] at (3.2,7.6) {\scriptsize $K_2$};
\node [rotate=0.] at (6.,9.2) {\scriptsize $K_3$};
\node [rotate=0.] at (3.7,7.6) {\scriptsize $\mathrm{PB}$};
\node [rotate=0.] at (3.5,9.8) {\scriptsize $\mathrm{PB}$};
\node [rotate=0.] at (6.5,9.2) {\scriptsize $\mathrm{PB}$};
\end{tikzpicture}
}

\newcommand{\GenericSurfaceConfig}{
\begin{tikzpicture}[baseline={([yshift=-.8ex]current bounding box.center)},scale=1]
\draw [lightgray, fill] (2.5,10.4) ..controls (1.71,10.193) and (1.654,9.7) .. (1.6,9.3)
			 ..controls (1.546,8.9) and (1.493,8.593) .. (1.6,8.2)
			 ..controls (1.707,7.807) and (1.972,7.329) .. (2.4,7.1)
			 ..controls (2.828,6.871) and (3.419,6.892) .. (4.1,6.9)
			 ..controls (4.781,6.908) and (5.552,6.903) .. (6.2,7.)
			 ..controls (6.848,7.097) and (7.372,7.297) .. (7.6,7.8)
			 ..controls (7.828,8.303) and (7.76,9.108) .. (7.5,9.6)
			 ..controls (7.24,10.092) and (6.788,10.271) .. (5.8,10.4)
			 ..controls (4.812,10.529) and (3.29,10.607) .. (2.5,10.4);
\draw [white, fill] (2.9,10.) ..controls (2.675,9.95) and (2.8,9.65) .. (2.9,9.5)
			 ..controls (3.,9.35) and (3.075,9.35) .. (3.3,9.4)
			 ..controls (3.525,9.45) and (3.9,9.55) .. (3.8,9.7)
			 ..controls (3.7,9.85) and (3.125,10.05) .. (2.9,10.);
\draw [white, fill] (3.1,7.9) ..controls (2.9,7.775) and (2.8,7.45) .. (3.,7.3)
			 ..controls (3.2,7.15) and (3.7,7.175) .. (3.9,7.3)
			 ..controls (4.1,7.425) and (4.,7.65) .. (3.8,7.8)
			 ..controls (3.6,7.95) and (3.3,8.025) .. (3.1,7.9);
\draw [white, fill] (4.1,8.7) ..controls (4.1,8.575) and (4.325,8.325) .. (4.5,8.3)
			 ..controls (4.675,8.275) and (4.8,8.475) .. (4.8,8.6)
			 ..controls (4.8,8.725) and (4.675,8.775) .. (4.5,8.8)
			 ..controls (4.325,8.825) and (4.1,8.825) .. (4.1,8.7);
\draw [white, fill] (5.7,9.3) ..controls (5.6,9.075) and (5.7,8.75) .. (6.,8.7)
			 ..controls (6.3,8.65) and (6.8,8.875) .. (6.9,9.1)
			 ..controls (7.,9.325) and (6.7,9.55) .. (6.4,9.6)
			 ..controls (6.1,9.65) and (5.8,9.525) .. (5.7,9.3);
\draw [] (2.5,10.4) ..controls (1.71,10.193) and (1.654,9.7) .. (1.6,9.3)
			 ..controls (1.546,8.9) and (1.493,8.593) .. (1.6,8.2)
			 ..controls (1.707,7.807) and (1.972,7.329) .. (2.4,7.1)
			 ..controls (2.828,6.871) and (3.419,6.892) .. (4.1,6.9)
			 ..controls (4.781,6.908) and (5.552,6.903) .. (6.2,7.)
			 ..controls (6.848,7.097) and (7.372,7.297) .. (7.6,7.8)
			 ..controls (7.828,8.303) and (7.76,9.108) .. (7.5,9.6)
			 ..controls (7.24,10.092) and (6.788,10.271) .. (5.8,10.4)
			 ..controls (4.812,10.529) and (3.29,10.607) .. (2.5,10.4);
\draw [] (2.9,10.) ..controls (2.675,9.95) and (2.8,9.65) .. (2.9,9.5)
			 ..controls (3.,9.35) and (3.075,9.35) .. (3.3,9.4)
			 ..controls (3.525,9.45) and (3.9,9.55) .. (3.8,9.7)
			 ..controls (3.7,9.85) and (3.125,10.05) .. (2.9,10.);
\draw [] (3.1,7.9) ..controls (2.9,7.775) and (2.8,7.45) .. (3.,7.3)
			 ..controls (3.2,7.15) and (3.7,7.175) .. (3.9,7.3)
			 ..controls (4.1,7.425) and (4.,7.65) .. (3.8,7.8)
			 ..controls (3.6,7.95) and (3.3,8.025) .. (3.1,7.9);
\draw [] (5.7,9.3) ..controls (5.6,9.075) and (5.7,8.75) .. (6.,8.7)
			 ..controls (6.3,8.65) and (6.8,8.875) .. (6.9,9.1)
			 ..controls (7.,9.325) and (6.7,9.55) .. (6.4,9.6)
			 ..controls (6.1,9.65) and (5.8,9.525) .. (5.7,9.3);
\draw [] (4.5,8.8) ..controls (4.325,8.825) and (4.1,8.825) .. (4.1,8.7)
			 ..controls (4.1,8.575) and (4.325,8.325) .. (4.5,8.3)
			 ..controls (4.675,8.275) and (4.8,8.475) .. (4.8,8.6)
			 ..controls (4.8,8.725) and (4.675,8.775) .. (4.5,8.8);
\draw [very thick] (2.9,9.5) ..controls (2.833,9.283) and (2.767,9.067) .. (2.8,8.8)
			 ..controls (2.833,8.533) and (2.967,8.217) .. (3.1,7.9);
\draw [very thick] (3.9,7.3) ..controls (4.475,7.283) and (5.05,7.267) .. (5.4,7.5)
			 ..controls (5.75,7.733) and (5.875,8.217) .. (6.,8.7);
\draw [very thick] (3.8,9.7) ..controls (4.242,9.683) and (4.683,9.667) .. (5.,9.6)
			 ..controls (5.317,9.533) and (5.508,9.417) .. (5.7,9.3);
\draw [very thick] (4.1,9.2) ..controls (3.836,9.1) and (3.727,8.918) .. (3.7,8.7)
			 ..controls (3.673,8.482) and (3.727,8.227) .. (4.,8.1)
			 ..controls (4.273,7.973) and (4.764,7.973) .. (5.,8.2)
			 ..controls (5.236,8.427) and (5.218,8.882) .. (5.,9.1)
			 ..controls (4.782,9.318) and (4.364,9.3) .. (4.1,9.2);
\node [rotate=0.] at (2.6,9.5) {\scriptsize $k_1$};
\node [rotate=0.] at (3.9,9.9) {\scriptsize $k_2$};
\node [rotate=0.] at (2.8,8.) {\scriptsize $k_3$};
\node [rotate=0.] at (4.1,7.1) {\scriptsize $k_4$};
\node [rotate=0.] at (6.2,8.5) {\scriptsize $k_5$};
\node [rotate=0.] at (5.6,9.6) {\scriptsize $k_6$};
\node [rotate=0.] at (3.,8.8) {\scriptsize $g_v$};
\node [rotate=0.] at (3.6,9.) {\scriptsize $g'_v$};
\node [rotate=0.] at (5.4,7.8) {\scriptsize $g_v$};
\node [rotate=0.] at (4.8,9.5) {\scriptsize $g_v$};
\end{tikzpicture}
}

\newcommand{\GSSphere}{
\begin{tikzpicture}[baseline={([yshift=-.8ex]current bounding box.center)},scale=1]
\draw [gray] (3.5,10.) ..controls (3.837,9.751) and (3.949,9.337) .. (4.,9.)
			 ..controls (4.051,8.663) and (4.041,8.405) .. (4.,8.1)
			 ..controls (3.959,7.795) and (3.885,7.444) .. (3.6,7.2)
			 ..controls (3.315,6.956) and (2.817,6.82) .. (2.4,7.)
			 ..controls (1.983,7.18) and (1.646,7.678) .. (1.6,8.3)
			 ..controls (1.554,8.922) and (1.798,9.668) .. (2.2,10.)
			 ..controls (2.602,10.332) and (3.163,10.249) .. (3.5,10.);
\draw [->] (2.4,8.6) ..controls (2.617,8.942) and (2.833,9.283) .. (3.,9.5)
			 ..controls (3.167,9.717) and (3.283,9.808) .. (3.4,9.9);
\draw [->] (2.4,8.6) ..controls (2.733,8.742) and (3.067,8.883) .. (3.3,9.)
			 ..controls (3.533,9.117) and (3.667,9.208) .. (3.8,9.3);
\draw [->] (2.4,8.6) ..controls (2.625,8.542) and (2.85,8.483) .. (3.1,8.5)
			 ..controls (3.35,8.517) and (3.625,8.608) .. (3.9,8.7);
\draw [->] (2.4,8.6) ..controls (2.625,8.425) and (2.85,8.25) .. (3.1,8.2)
			 ..controls (3.35,8.15) and (3.625,8.225) .. (3.9,8.3);
\draw [->] (2.4,8.6) ..controls (2.533,8.275) and (2.667,7.95) .. (2.9,7.8)
			 ..controls (3.133,7.65) and (3.467,7.675) .. (3.8,7.7);
\draw [->] (5.8,9.4) ..controls (5.183,9.742) and (4.567,10.083) .. (4.2,10.2)
			 ..controls (3.833,10.317) and (3.717,10.208) .. (3.6,10.1);
\draw [->] (5.8,9.4) ..controls (5.56,9.32) and (5.32,9.24) .. (5.1,9.3)
			 ..controls (4.88,9.36) and (4.68,9.56) .. (4.5,9.6)
			 ..controls (4.32,9.64) and (4.16,9.52) .. (4.,9.4);
\draw [->] (5.8,9.4) ..controls (5.589,9.28) and (5.378,9.16) .. (5.2,9.1)
			 ..controls (5.022,9.04) and (4.878,9.04) .. (4.7,9.)
			 ..controls (4.522,8.96) and (4.311,8.88) .. (4.1,8.8);
\draw [->] (5.8,9.4) ..controls (5.762,9.176) and (5.724,8.951) .. (5.6,8.8)
			 ..controls (5.476,8.649) and (5.264,8.571) .. (5.,8.5)
			 ..controls (4.736,8.429) and (4.418,8.364) .. (4.1,8.3);
\draw [->] (5.8,9.4) ..controls (5.853,9.016) and (5.907,8.631) .. (5.8,8.4)
			 ..controls (5.693,8.169) and (5.427,8.091) .. (5.1,8.)
			 ..controls (4.773,7.909) and (4.387,7.804) .. (4.,7.7);
\draw [fill,rotate around={0.:(2.4,8.6)}] (2.4,8.6) ++(0.:0.08 and 0.08) arc(0.:360.:0.08 and 0.08);
\draw [fill,rotate around={0.:(5.8,9.4)}] (5.8,9.4) ++(0.:0.08 and 0.08) arc(0.:360.:0.08 and 0.08);
\node [rotate=0.] at (2.3,8.8) {\scriptsize $v_A$};
\node [rotate=0.] at (5.9,9.7) {\scriptsize $v_B$};
\node [rotate=0.] at (3.1,10.) {\scriptsize $g_{1}$};
\node [rotate=0.] at (3.6,9.4) {\scriptsize $g_{2}$};
\node [rotate=0.] at (3.8,8.9) {\scriptsize $g_{3}$};
\node [rotate=0.] at (3.6,10.4) {\scriptsize $g'_{1}$};
\node [rotate=0.] at (4.,9.7) {\scriptsize $g'_{2}$};
\node [rotate=0.] at (4.2,9.1) {\scriptsize $g'_{3}$};
\end{tikzpicture}
}

\newcommand{\LeftEdge}[1]{
\begin{tikzpicture}[baseline={([yshift=-.8ex]current bounding box.center)},scale=1]
\draw [->-] (2.8,10.2) -- (1.8,10.2);
\draw [] (1.8,10.4) -- (1.8,10.);
\draw [] (2.8,10.4) -- (2.8,10.);
\draw [] (2.8,10.2) -- (3.,10.2);
\draw [] (1.6,10.2) -- (1.8,10.2);
\node [rotate=0.] at (2.3,10.4) {\scriptsize $#1$};
\end{tikzpicture}
}

\newcommand{\NonLocalTransform}{
\begin{tikzpicture}[baseline={([yshift=-.8ex]current bounding box.center)},scale=1]
\draw [->-] (1.9,8.6) -- (2.7,8.6);
\draw [->-] (2.7,8.6) -- (2.7,9.4);
\draw [->-] (2.7,9.4) -- (2.7,10.2);
\draw [->-] (1.9,8.6) -- (1.9,9.4);
\draw [->-] (1.9,9.4) -- (1.9,10.2);
\draw [->-] (1.9,10.2) -- (2.7,10.2);
\draw [->-] (1.9,9.4) -- (2.7,9.4);
\node [rotate=0.] at (2.3,8.4) {\scriptsize $a_{1}$};
\node [rotate=0.] at (3.,9.) {\scriptsize $a_{2}$};
\node [rotate=0.] at (1.6,9.) {\scriptsize $a_{3}$};
\node [rotate=0.] at (2.3,9.6) {\scriptsize $a_{4}$};
\node [rotate=0.] at (1.6,9.9) {\scriptsize $a_{5}$};
\node [rotate=0.] at (3.,9.9) {\scriptsize $a_{6}$};
\node [rotate=0.] at (2.3,10.4) {\scriptsize $a_{7}$};
\end{tikzpicture}
}

\newcommand{\NonLocalTransformAA}{
\begin{tikzpicture}[baseline={([yshift=-.8ex]current bounding box.center)},scale=1]
\draw [->] (2.,8.8) ..controls (1.958,8.933) and (1.917,9.067) .. (1.9,9.2)
			 ..controls (1.883,9.333) and (1.892,9.467) .. (1.9,9.6);
\draw [->] (2.,8.8) ..controls (2.008,9.067) and (2.017,9.333) .. (2.,9.6)
			 ..controls (1.983,9.867) and (1.942,10.133) .. (1.9,10.4);
\draw [->] (2.,8.8) ..controls (2.052,8.969) and (2.105,9.138) .. (2.1,9.4)
			 ..controls (2.095,9.662) and (2.033,10.017) .. (2.,10.2)
			 ..controls (1.967,10.383) and (1.962,10.395) .. (2.1,10.4)
			 ..controls (2.238,10.405) and (2.519,10.402) .. (2.8,10.4);
\draw [->] (2.,8.8) ..controls (2.124,8.858) and (2.249,8.916) .. (2.4,8.9)
			 ..controls (2.551,8.884) and (2.729,8.796) .. (2.8,8.9)
			 ..controls (2.871,9.004) and (2.836,9.302) .. (2.8,9.6);
\draw [->] (2.,8.8) -- (2.8,8.8);
\draw [fill,rotate around={0.:(2.,8.8)}] (2.,8.8) ++(0.:0.08 and 0.08) arc(0.:360.:0.08 and 0.08);
\node [rotate=0.] at (2.8,8.6) {\scriptsize $g_{1}$};
\node [rotate=0.] at (3.,9.6) {\scriptsize $g_{2}$};
\node [rotate=0.] at (1.7,9.6) {\scriptsize $g_{3}$};
\node [rotate=0.] at (1.6,10.4) {\scriptsize $g_{4}$};
\node [rotate=0.] at (2.8,10.2) {\scriptsize $g_{5}$};
\end{tikzpicture}
}

\newcommand{\PBEB}{
\begin{tikzpicture}[baseline={([yshift=-.8ex]current bounding box.center)},scale=1]
\draw [lightgray, fill] (2.6,10.8) ..controls (2.305,10.741) and (2.3,10.562) .. (2.2,10.2)
			 ..controls (2.1,9.837) and (1.905,9.291) .. (2.2,9.1)
			 ..controls (2.495,8.909) and (3.279,9.073) .. (4.,9.1)
			 ..controls (4.721,9.127) and (5.38,9.016) .. (5.7,9.2)
			 ..controls (6.02,9.384) and (6.,9.862) .. (5.9,10.2)
			 ..controls (5.8,10.538) and (5.62,10.734) .. (5.3,10.8)
			 ..controls (4.98,10.866) and (4.521,10.802) .. (4.,10.8)
			 ..controls (3.479,10.798) and (2.895,10.859) .. (2.6,10.8);
\draw [] (2.6,10.8) ..controls (2.305,10.741) and (2.3,10.562) .. (2.2,10.2)
			 ..controls (2.1,9.837) and (1.905,9.291) .. (2.2,9.1)
			 ..controls (2.495,8.909) and (3.279,9.073) .. (4.,9.1)
			 ..controls (4.721,9.127) and (5.38,9.016) .. (5.7,9.2)
			 ..controls (6.02,9.384) and (6.,9.862) .. (5.9,10.2)
			 ..controls (5.8,10.538) and (5.62,10.734) .. (5.3,10.8)
			 ..controls (4.98,10.866) and (4.521,10.802) .. (4.,10.8)
			 ..controls (3.479,10.798) and (2.895,10.859) .. (2.6,10.8);
\draw [->] (3.,10.9) ..controls (3.086,10.974) and (3.171,11.048) .. (3.6,11.1)
			 ..controls (4.029,11.152) and (4.8,11.183) .. (5.3,11.1)
			 ..controls (5.8,11.017) and (6.029,10.819) .. (6.2,10.7)
			 ..controls (6.371,10.581) and (6.486,10.54) .. (6.6,10.5);
\draw [thick] (4.,9.1) -- (4.,10.8);
\draw [->] (5.9,10.4) -- (6.6,10.4);
\draw [->] (4.1,9.8) -- (6.6,9.8);
\node [rotate=0.] at (8.4,10.5) {\scriptsize $\mathrm{Physical Boundary (PB)}$};
\node [rotate=0.] at (8.8,9.8) {\scriptsize $\mathrm{Entanglement Boundary (EB)}$};
\node [rotate=0.] at (3.,10.2) {\scriptsize $A$};
\node [rotate=0.] at (5.,10.2) {\scriptsize $B$};
\end{tikzpicture}
}

\newcommand{\PlaqquetteState}[4]{
\begin{tikzpicture}[baseline={([yshift=-.8ex]current bounding box.center)},scale=1]
\draw [->-] (1.9,10.2) -- (1.9,9.2);
\draw [->-] (1.9,9.2) -- (2.9,9.2);
\draw [->-] (2.9,9.2) -- (2.9,10.2);
\draw [->-] (2.9,10.2) -- (1.9,10.2);
\draw [] (1.7,10.2) -- (1.9,10.2);
\draw [] (1.9,10.2) -- (1.9,10.4);
\draw [] (2.9,10.2) -- (3.1,10.2);
\draw [] (2.9,10.4) -- (2.9,10.2);
\draw [] (1.7,9.2) -- (1.9,9.2);
\draw [] (1.9,9.2) -- (1.9,9.);
\draw [] (2.9,9.) -- (2.9,9.2);
\draw [] (2.9,9.2) -- (3.1,9.2);
\node [rotate=0.] at (1.6,9.7) {\scriptsize $#1$};
\node [rotate=0.] at (2.4,10.4) {\scriptsize $#2$};
\node [rotate=0.] at (3.1,9.7) {\scriptsize $#3$};
\node [rotate=0.] at (2.4,9.) {\scriptsize $#4$};
\end{tikzpicture}
}

\newcommand{\RightEdge}[1]{
\begin{tikzpicture}[baseline={([yshift=-.8ex]current bounding box.center)},scale=1]
\draw [->-] (1.8,10.2) -- (2.8,10.2);
\draw [] (1.8,10.4) -- (1.8,10.);
\draw [] (2.8,10.4) -- (2.8,10.);
\draw [] (2.8,10.2) -- (3.,10.2);
\draw [] (1.6,10.2) -- (1.8,10.2);
\node [rotate=0.] at (2.3,10.4) {\scriptsize $#1$};
\end{tikzpicture}
}

\newcommand{\SchmidtDecomp}{
\begin{tikzpicture}[baseline={([yshift=-.8ex]current bounding box.center)},scale=1]
\draw [] (3.7,10.) ..controls (3.712,10.336) and (3.737,10.379) .. (3.4,10.4)
			 ..controls (3.062,10.421) and (2.362,10.421) .. (2.,10.4)
			 ..controls (1.637,10.379) and (1.612,10.336) .. (1.6,10.)
			 ..controls (1.587,9.664) and (1.587,9.036) .. (1.6,8.7)
			 ..controls (1.612,8.364) and (1.638,8.321) .. (2.,8.3)
			 ..controls (2.362,8.279) and (3.062,8.279) .. (3.4,8.3)
			 ..controls (3.738,8.321) and (3.712,8.364) .. (3.7,8.7)
			 ..controls (3.688,9.036) and (3.688,9.664) .. (3.7,10.);
\draw [] (6.,10.) ..controls (5.987,10.336) and (5.962,10.379) .. (5.6,10.4)
			 ..controls (5.237,10.421) and (4.537,10.421) .. (4.2,10.4)
			 ..controls (3.862,10.379) and (3.887,10.336) .. (3.9,10.)
			 ..controls (3.912,9.664) and (3.912,9.036) .. (3.9,8.7)
			 ..controls (3.887,8.364) and (3.862,8.321) .. (4.2,8.3)
			 ..controls (4.538,8.279) and (5.238,8.279) .. (5.6,8.3)
			 ..controls (5.962,8.321) and (5.987,8.364) .. (6.,8.7)
			 ..controls (6.012,9.036) and (6.012,9.664) .. (6.,10.);
\draw [gray,fill,rotate around={0.:(3.7,9.8)}] (3.7,9.8) ++(0.:0.08 and 0.08) arc(0.:360.:0.08 and 0.08);
\draw [gray,fill,rotate around={0.:(3.7,9.4)}] (3.7,9.4) ++(0.:0.08 and 0.08) arc(0.:360.:0.08 and 0.08);
\draw [gray,fill,rotate around={0.:(3.7,9.3)}] (3.7,9.3) ++(0.:0.08 and 0.08) arc(0.:360.:0.08 and 0.08);
\draw [gray,fill,rotate around={0.:(3.7,9.1)}] (3.7,9.1) ++(0.:0.08 and 0.08) arc(0.:360.:0.08 and 0.08);
\draw [gray,fill,rotate around={0.:(3.7,8.5)}] (3.7,8.5) ++(0.:0.08 and 0.08) arc(0.:360.:0.08 and 0.08);
\draw [gray,fill,rotate around={0.:(3.9,9.8)}] (3.9,9.8) ++(0.:0.08 and 0.08) arc(0.:360.:0.08 and 0.08);
\draw [gray,fill,rotate around={0.:(3.9,9.4)}] (3.9,9.4) ++(0.:0.08 and 0.08) arc(0.:360.:0.08 and 0.08);
\draw [gray,fill,rotate around={0.:(3.9,9.3)}] (3.9,9.3) ++(0.:0.08 and 0.08) arc(0.:360.:0.08 and 0.08);
\draw [gray,fill,rotate around={0.:(3.9,9.1)}] (3.9,9.1) ++(0.:0.08 and 0.08) arc(0.:360.:0.08 and 0.08);
\draw [gray,fill,rotate around={0.:(3.9,8.5)}] (3.9,8.5) ++(0.:0.08 and 0.08) arc(0.:360.:0.08 and 0.08);
\draw [gray,fill,rotate around={0.:(3.7,10.2)}] (3.7,10.2) ++(0.:0.08 and 0.08) arc(0.:360.:0.08 and 0.08);
\draw [gray,fill,rotate around={0.:(3.9,10.2)}] (3.9,10.2) ++(0.:0.08 and 0.08) arc(0.:360.:0.08 and 0.08);
\node [rotate=0.] at (2.4,9.1) {\scriptsize $A$};
\node [rotate=0.] at (5.3,9.1) {\scriptsize $B$};
\node [rotate=0.] at (3.3,9.6) {\scriptsize $\ket{\Psi_i}_{\! A}$};
\node [rotate=0.] at (4.3,9.6) {\scriptsize $\ket{\Psi_i}_{\! B}$};
\end{tikzpicture}
}

\newcommand{\torusChargeBasis}{
\begin{tikzpicture}[baseline={([yshift=-.8ex]current bounding box.center)},scale=1]
\draw [gray] (3.4,9.2) ..controls (3.6,9.3) and (3.8,9.4) .. (4.,9.4)
			 ..controls (4.2,9.4) and (4.4,9.3) .. (4.6,9.2);
\draw [gray] (3.2,9.4) ..controls (3.233,9.343) and (3.267,9.286) .. (3.4,9.2)
			 ..controls (3.533,9.114) and (3.767,9.) .. (4.,9.)
			 ..controls (4.233,9.) and (4.467,9.114) .. (4.6,9.2)
			 ..controls (4.733,9.286) and (4.767,9.343) .. (4.8,9.4);
\draw [] (2.8,8.) ..controls (2.745,8.079) and (2.69,8.158) .. (2.7,8.3)
			 ..controls (2.71,8.442) and (2.783,8.646) .. (2.9,8.8)
			 ..controls (3.017,8.954) and (3.176,9.058) .. (3.3,9.1)
			 ..controls (3.424,9.142) and (3.512,9.121) .. (3.6,9.1);
\draw [dashed] (3.6,9.1) ..controls (3.644,9.06) and (3.688,9.02) .. (3.7,8.9)
			 ..controls (3.712,8.78) and (3.692,8.579) .. (3.6,8.4)
			 ..controls (3.508,8.221) and (3.345,8.063) .. (3.2,8.)
			 ..controls (3.055,7.937) and (2.927,7.968) .. (2.8,8.);
\draw [] (2.9,8.8) ..controls (3.057,8.695) and (3.215,8.59) .. (3.5,8.5)
			 ..controls (3.785,8.41) and (4.198,8.335) .. (4.6,8.4)
			 ..controls (5.002,8.465) and (5.394,8.671) .. (5.6,8.9)
			 ..controls (5.806,9.129) and (5.828,9.381) .. (5.7,9.6)
			 ..controls (5.572,9.819) and (5.296,10.003) .. (5.,10.1)
			 ..controls (4.704,10.197) and (4.39,10.205) .. (4.1,10.2)
			 ..controls (3.81,10.195) and (3.546,10.176) .. (3.3,10.1)
			 ..controls (3.054,10.024) and (2.826,9.891) .. (2.7,9.7)
			 ..controls (2.574,9.509) and (2.55,9.26) .. (2.6,9.1)
			 ..controls (2.65,8.94) and (2.775,8.87) .. (2.9,8.8);
\draw [gray] (1.6,9.2) ..controls (1.6,9.68) and (2.08,10.16) .. (2.8,10.4)
			 ..controls (3.52,10.64) and (4.48,10.64) .. (5.2,10.4)
			 ..controls (5.92,10.16) and (6.4,9.68) .. (6.4,9.2)
			 ..controls (6.4,8.72) and (5.92,8.24) .. (5.2,8.)
			 ..controls (4.48,7.76) and (3.52,7.76) .. (2.8,8.)
			 ..controls (2.08,8.24) and (1.6,8.72) .. (1.6,9.2);
\draw [->] (2.9,8.8) ..controls (3.525,8.85) and (4.15,8.9) .. (4.5,9.)
			 ..controls (4.85,9.1) and (4.925,9.25) .. (5.,9.4);
\draw [->] (2.9,8.8) ..controls (2.933,8.927) and (2.967,9.053) .. (3.,9.2)
			 ..controls (3.033,9.347) and (3.067,9.513) .. (3.2,9.6)
			 ..controls (3.333,9.687) and (3.567,9.693) .. (3.8,9.7);
\draw [->] (2.9,8.8) ..controls (3.222,8.804) and (3.544,8.809) .. (3.8,8.8)
			 ..controls (4.056,8.791) and (4.244,8.769) .. (4.4,8.8)
			 ..controls (4.556,8.831) and (4.678,8.916) .. (4.8,9.);
\draw [->] (2.9,8.8) ..controls (3.343,8.709) and (3.786,8.618) .. (4.1,8.6)
			 ..controls (4.414,8.582) and (4.6,8.638) .. (4.8,8.8)
			 ..controls (5.,8.962) and (5.214,9.232) .. (5.2,9.4)
			 ..controls (5.186,9.568) and (4.943,9.634) .. (4.7,9.7);
\draw [->] (2.9,8.8) ..controls (2.758,8.783) and (2.617,8.767) .. (2.5,8.8)
			 ..controls (2.383,8.833) and (2.292,8.917) .. (2.2,9.);
\draw [->] (2.9,8.8) ..controls (2.698,8.731) and (2.496,8.662) .. (2.3,8.7)
			 ..controls (2.104,8.738) and (1.916,8.882) .. (1.9,9.1)
			 ..controls (1.884,9.318) and (2.042,9.609) .. (2.2,9.9);
\draw [fill,rotate around={0.:(2.9,8.8)}] (2.9,8.8) ++(0.:0.08 and 0.08) arc(0.:360.:0.08 and 0.08);
\node [rotate=0.] at (5.6,8.6) {\scriptsize $C_1$};
\node [rotate=0.] at (3.,8.2) {\scriptsize $C_2$};
\end{tikzpicture}
}
\title{Entanglement Entropy, Quantum Fluctuations, and Thermal Entropy in Topological Phases}
\date{\today}
\author[d,e]{Yuting Hu}
\author[a,b,c,d,f]{Yidun Wan}
\affiliation[a]{State Key Laboratory of Surface Physics, Fudan University, Shanghai 200433, China}
\affiliation[b]{Department of Physics and Center for Field Theory and Particle Physics, Fudan University, Shanghai 200433, China}
\affiliation[c]{Institute for Nanoelectronic devices and Quantum computing, Fudan University, Shanghai 200433, China}
\affiliation[d]{Department of Physics and Institute for Quantum Science and Engineering, Southern University of Science and Technology, Shenzhen 518055, China}
\affiliation[e]{CAS Key Laboratory of Microscale Magnetic Resonance and Department of Modern Physics, University of Science and Technology of China, Hefei,
Anhui 230026, China}
\affiliation[f]{Collaborative Innovation Center of Advanced Microstructures, Nanjing, 210093, China}
\emailAdd{yuting.phys@gmail.com, ydwan@fudan.edu.cn}

\abstract{
Entanglement entropy in topologically ordered matter phases has been computed extensively using various methods. In this paper, we study the entanglement entropy of topological phases in two-spaces from a new perspective---the perspective of quasiparticle fluctuations. In this picture, the entanglement spectrum of a topologically ordered system is identified with the spectrum of quasiparticle fluctuations of the system, and the entanglement entropy measures the maximal quasiparticle fluctuations on the EB. As a consequence, entanglement entropy corresponds to the thermal entropy of the quasiparticles at infinite temperature on the entanglement boundary. We corroborates our results with explicit computation in the quantum double model with/without boundaries. We then systematically construct the reduced density matrices of the quantum double model on generic 2-surfaces with boundaries. 
}

\begin{document}

\maketitle
\flushbottom

\section{Introduction}

Matter phases with intrinsic topological orders, or topological orders for short, are exotic, gapped phases of matter beyond the Landau-Ginzburg paradigm\cite{Wen1989,Wen1989a,Wen1990a,Wen1990c,Kitaev2003a,Levin2004,Kitaev2006}. They not only have expanded our knowledge of phases of matter but also have important applications, including robust quantum memories and topological quantum computers\cite{Kitaev2003a,Nayak2008}.  

Topological orders exhibit long range entanglement\cite{Chen2010e,Wen2016}, which can be captured by the topological entanglement entropy (TEE)\cite{Levin2006,Kitaev2006} of the system. Suppose a $(2+1)$-dimensional topologically ordered system is virtually divided into two regions $A$ and $B$, the entanglement entropy (EE) of the system is the sum of an leading area term, proportional to the length of the boundary between the two regions, and a universal subleading term proportional to $\log D$, where $D$ is the total quantum dimension\footnote{For a topologically ordered system with $n$ type of anyons and $d_i$ is the quantum dimension of the type-$i$ anyons, $D=\sqrt{\sum_{i=1}^n d_i^2}$. }. As such, TEE can be used to distinguish topological orders unless certain topological orders has the same total quantum dimension.

\begin{figure}[!ht]
\centering
\includegraphics[scale=0.45]{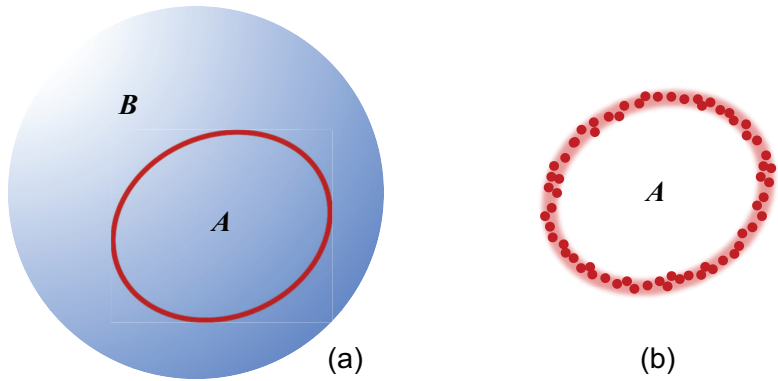}
\caption{(a) is a system divided into regions $A$ and $B$ by a circular EB (in red). Tracing out region $B$ in (a) is equivalent to (b), where there is only region $A$ with a boundary decorated by quasiparticle fluctuation. The EE between $A$ and $B$ in (a) is equal to the TE of the quasiparticles on the boundary of $A$ in (b). }
\label{fig:TE=EE}
\end{figure}

There has been extensive works computing the EE in topological orders\cite{Levin2006,Kitaev2006b,Dong2008,Flammia2009,Yao2010a,Zhang2011,Grover2011,Brown2013,HungWan2015,Luo2016b,Luo2018,Chen2018,Wen2018,Shi2018,Shi2019}. In this paper, however, we propose a novel perspective of the EE in topological orders in two spatial dimensions, namely, the picture of \emph{quasiparticle fluctuations}.  Consider a bipartite system in a ground-state, e.g., as in Fig. \ref{fig:TE=EE}(a). We show as one of our main results that the reduced density matrix $\rho_A$ is identified with the quasiparticle fluctuations on the entanglement boundary (EB) separating $A$ and $B$, and the corresponding EE explicitly counts the allowed states describing quasiparticle fluctuations on the EB. We shall call such states EB-states. An EB-state is allowed when it ensures that no quasiparticles exist anywhere in the system except on the EB and that the total charge of the system and that of the EB are trivial. Hence, these quasiparticles cannot be seen by any global observer, say \emph{Gabe}, of the system, as each of them is confined in a pair with its anti-quasiparticle on the other side of the EB. Such quasiparticle-antiquasiparticle pairs, which happen to be cut by the EB, comprise the quasiparticle fluctuations along the EB. We reckon that our notion of quasiparticle fluctuations complies with the usual notion of vacuum fluctuations in quantum field theory, which refers to that the vacuum consists of instantaneous particle-antiparticle pairs everywhere. Particles in vacuum fluctuations cannot be observed unless the particle-antiparticle pairs are torn apart locally by sufficient amount of energy.

Entanglement entropy counts the EB-states because of the following. To compute the EE, Gabe has to trace out region $B$. Tracing out region $B$ would blind Alice from knowing the existence of region $B$ at all. What Alice would see is a world like that in Fig. \ref{fig:TE=EE}(b), complete darkness (no quasiparticles in the bulk) with a glowing physical boundary (quasiparticles residing on the boundary).  Alice will detect quasiparticles at the EB, which now appears to her as a physical boundary.

Furthermore, Alice is able to measure the EB-states on the EB and find that they are all equally probable because the reduced density matrix $\rho_A$ can be made a projector. She can then compute the thermal entropy (TE) of the boundary quasiparticle system at temperature $T=\infty$. This TE would coincide with the EE computed by Gabe. The infinite temperature can be thought as what offers the energy to tear apart the particle-antiparticle pairs at the EB.

To make the above picture of quantum fluctuations concrete and precise, we consider in particular the quantum double (QD) model of topological orders in two-spaces and focus on the ground-state Hilbert space of the model. The QD model\cite{Kitaev2006,Hu2012a,Mesaros2011} is a lattice Hamiltonian extension of the Dijkgraaf-Witten gauge theory\cite{Dijkgraaf1990} with discrete gauge groups. We focus on ground-states because anyon excitations in topological orders add no more physical content to the EE. 

In the QD model, we construct two useful types of bases---the holonomy bases and the fusion bases.
The holonomy bases and fusion bases not only simplify the EE computation but also manifest the picture of quasiparticle fluctuations of EE. We compute the EE on a sphere in both bases, which leads to the same result, which agrees with the known result. This result corroborates our picture of quasiparticle fluctuations.

\begin{figure}[h]
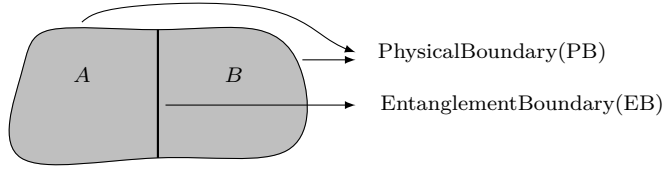

        \centering
        \PBEB
        \caption{Physical boundary (PB) and entanglement boundary (EB).}
        \label{fig:PBEB}
\end{figure}

We then extend our study to topological orders with gapped physical boundaries (PBs), separating the system from the vacuum, as opposed to the EB separating two regions of the system. See Fig. \ref{fig:PBEB}.
We consider the extended QD model\cite{Beigi2011,Bullivant2017}, dedicated to the scenario with PBs. With minor modifications, the holonomy and fusion bases work in this scenario, and hence the picture of quasiparticle fluctuations still applies. To be specific, we considered two representative cases, namely, a cylinder with region $A$ being a meridian strip and a cylinder with region $A$ being a longitudinal strip.\footnote{The directions are defined as in the Mercator projection.} Using the holonomy and fusion bases, we offer closed-form formulae for the EE in both cases and an explicit example when the bulk gauge group is the permutation group $S_3$. Such a formula in terms of the input data of the QD model has not been obtained before. We then systematically construct the reduced density matrices for topological orders on generic two-surfaces with gapped PBs.

We note that the EE on topological orders with PBs have also been computed recently on the extended QD model\cite{Chen2018} and by using conformal field theory techniques\cite{Wen2016a,Hung2019}. 

We reckon that the fusion bases we construct are related to the fusion bases used in a recent work studying the generalized anyon exclusion statistics\cite{Hu2013,Li2018}. Hence, we expect to relate the EE in topological orders to the generalized anyon exclusion statistics. This expectation is reasonable because exclusion statistics is bound to the notion of thermal entropy. We shall leave this connection for future studies.

This paper is organized as follows. Section \ref{sec:QD} reviews the QD model and develops the holonomy bases. Section \ref{sec:qpPicture} constructs the fusion bases and explains the picture of quasiparticle fluctuations. Section \ref{sec:Cyl} studies the cases with PBs, with a concrete example for $G=S_3$. Section \ref{sec:disc} discusses certain subtleties. The appendices collect certain calculations too detailed to be part of the main text.

\section{Quantum double model and holonomy bases}
\label{sec:QD}

In this section, we shall discuss the minimal and complete set of observables in a lattice topological gauge field theory. We will also discuss the relation between the gauge fields and these observables. This relation will enable us to study the entanglement from the perspective of observables. 

We define a discrete gauge field theory on a 2D directed graph, with a finite gauge group $G$. The degrees of freedom (d.o.f.) are group elements on the edges. The Hilbert space is spanned by all possible configurations of these group elements on the edges. We identify the two states,
\begin{equation}\label{eq:identityLR}
\bket{\LeftEdge{g}}\equiv \bket{\RightEdge{g^{-1}}},
\end{equation}
where on the RHS the group element is inversed and the arrow on the edge is reversed. For simplicity and illustration, we show only four-valent vertices but  in general the spatial graph could contain any multivalent vertex.

Then we define a gauge transformation at a vertex $v$ by
\begin{equation}\label{eq:GaugeTransform}
A_v^{h}\bket{\FourValentState{a}{b}{c}{d}}=\bket{\FourValentState{ha}{hb}{hc}{hd}},
\end{equation}
which is a left multiplication of $h$ on the edges coming into the vertex $v$.

A lattice gauge theory is defined by a Hamiltonian that commutes with any gauge transformation on the lattice.
The quantum double (QD) model, also known as the Kitaev model\cite{Kitaev2006}, is such a lattice gauge theory, whose Hamiltonian reads
\begin{equation}\label{eq:KitaevQDHamiltonian}
H=-\sum_vA_v-\sum_pB_p,
\end{equation}
where $A_v$ is a projection operator
\begin{equation}\label{eq:KitaevAv}
A_v=\frac{1}{|G|}\sum_{g\in G}A_v^g,
\end{equation}
which projects onto states with gauge invariance at vertex $v$. The other operator $B_p$ is an projection operator
\begin{equation}\label{eq:KitaevBp}
B_p\bket{\PlaqquetteState{a}{b}{c}{d}}=\delta_{abcd}\bket{\PlaqquetteState{a}{b}{c}{d}},
\end{equation}
where $\delta_{abcd}=1$ if $abcd=1$, the unit element of $G$, and $\delta_{abcd}=0$ otherwise. Here $abcd$ is the \emph{holonomy} around the plaquette $p$. Hence, $B_p$ projects onto states with trivial holonomy at $p$.

In particular, when $G=\Z_2$ (known as the toric code model), the Hilbert space is spanned by spins on the edges, and the Hamiltonian terms are
\begin{equation}\label{eq:KitaevTCAv}
A_v=\prod_{e\text{ into }v}\sigma^x_e,\quad B_p=\prod_{e\text{ around }p}\sigma^z_{e},
\end{equation}
where $e$ represent the edges. Indeed, compared to Eq. \eqref{eq:GaugeTransform}, $\prod_{e\text{ into }v}\sigma^x_e$ is (the non-identity) $\Z_2$ gauge transformation with the $\Z_2$ represented by $\{1,\sigma^x\}$. The delta function in Eq. \eqref{eq:KitaevBp} becomes $\frac{1}{2}\left(1+\prod_{e\text{ around }p}\sigma^z_{e}\right)$.

\subsection{Observables}

Since all the operators $A_v$ and $B_p$ commute with each other, they form commuting observables of the toric code model, and their common eigenvectors form the basis states of the Hamiltonian. As a results, there is a $\Z_2$ charge excitations at a vertex $v$ in a state $\ket{\psi}$ if $A_v\ket{\psi}=-\ket{\psi}$, and a $\Z_2$ flux excitation at a plaquette $p$ if $B_p\ket{\psi}=-\ket{\psi}$. The name ``charge'' and ``flux'' can be justified by the Aharonov-Bohm effect as follows. 

For a given edge $e$, the $\sigma^z_e$ serves as the hopping operator that moves the charge from one end $v_1$ of $e$ to the other end $v_2$ because $A_{v_{1,2}}\sigma^z_e\ket{\psi}=-\sigma^z_eA_{v_{1,2}}\ket{\psi}$. Hence, being the tensor product of $\sigma^z$ around the plaquette $p$, $B_p$ moves a charge around $p$ by one turn and measures the $\Z_2$ flux in the plaquette $p$. The names ``charge'' and ``flux'' thus follow.

On a torus, The minimal and complete set of commuting observables (also called the set of stabilizers) of the toric code model is given by
\begin{equation}\label{eq:setObservablesTC}
A_1,A_2,\dots,A_{V-1},B_1,B_2,\dots,B_{P-1},Z_1,Z_2,
\end{equation}
where $V$ is the total number of vertices, and $P$ the total number of plaquettes. The two operators $A_V$ and $B_P$ are excluded from the above set because of the the two global constraints
\begin{equation}\label{eq:globalConstraint}
\prod_{v=1}^V A_v=1,\quad \prod_{p=1}^P B_p=1.
\end{equation}
The  operators $Z_1$ and $Z_2$ are the products of $\sigma^z$ along two arbitrary, non-contractible loops, one meridian and one longitudinal, of the torus. These two operators clearly measure the $\Z_2$ fluxes going through the two holes of the torus (as seen in 3D space); they are responsible for the corresponding Aharonov-Bohm phases.

We arrive at two bases of the Hilbert space. One is spanned by the spins on all the edges, while the other is given by the simultaneous eigenvectors of the operators in Eq. \eqref{eq:setObservablesTC}. The latter corresponds to three types of physical quantities in a electromagnetic theory: $Z_2$ electric field strengths, magnetic field strengths, and Aharanov-Bohm phases due to nontrivial spatial topology. This establishes a duality between the observables and the gauge fields. We call the Aharonov-Bohm phases the topological d.o.f..

\subsection{Holonomy bases: a non-local transformation}\label{subsec:holoBasis}

The above duality between the gauge fields and observables can be also made precise in the QD model of a generic group $G$. For our purposes in this paper, we focus on the states with trivial holonomy, i.e., $B_p=1$, everywhere. That is, we restrict ourselves to the subspace $\Hil^{B_p=1}$ of the total Hilbert space of the QD model. 

\begin{figure}[h]
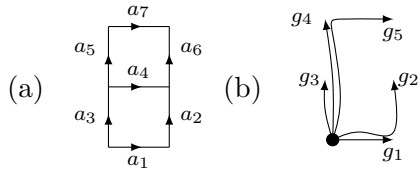

        \centering
(a) \NonLocalTransform 
(b) \NonLocalTransformAA
        \caption{(a) is a configuration of group elements on the edges of a graph, where trivial holonomy conditions $a_4a_3=a_2a_1$ and $a_7a_5=a_6a_4$ are assumed. (b) shows a non local transformation. The big dot (the lower-left vertex in (a)) is chosen as the base point. A path is chosen from the base point to every other vertex $v$ in (a). These paths are graced with the new degrees of freedom defined in Eq. \eqref{eq:pathGelement}. The explicit transformation consists of $g_1=a_1$, $g_2=a_2a_1$, $g_3=a_3$,$g_4=a_5a_3$, and $g_5=a_7a_5a_3$. Such a transformation from the $a$'s to $g$'s is clearly invertible.}
        \label{fig:nonLocalTransformExample}
\end{figure}

We define a non-local transformation as follows. First, we pick an arbitrary vertex $v_0$ as the base point and choose a path on the graph from $v_0$ to $v$, for any other vertex $v$. Second, Given a configuration of the group elements $a_e$ on all the edges satisfying the trivial holonomy condition, we assign to each vertex $v\neq v_0$ a new group element 
\be\label{eq:pathGelement}
g_v:=\prod_{e\in p_{vv_0}} a_e,
\ee
where $p_{vv_0}$ is the path chosen from $v$ to $v_0$. If the space has a non-contractible loop $C_\alpha$, we also assign a $g_{\alpha}$---the product of the group elements along a $C_{\alpha}$ to $C_\alpha$. This way, we get a non-local transformation: 
\begin{equation}\label{eq:nonLocalTransform}
\{a_e\} \cong \{g_v; g_{\alpha}\}_{v\neq v_0}.
\end{equation}
Such a transformation is illustrated in Fig. \ref{fig:nonLocalTransformExample}.

\begin{figure}[h]
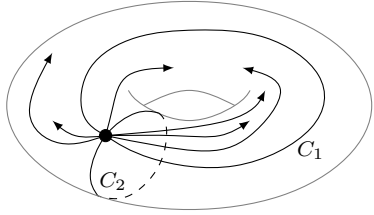

\centering
\torusChargeBasis
\caption{A basis of Hilbert space on torus, labeled by eigenvalues of $A_v$, ($v=1,2,\dots,V-1$), and $Z_1$, $Z_2$ defined along the two non-contractible loops $C_1$, $C_2.$. There is a base vertex shown as dot in the diagram. The gauge charge at vertex each vertex $v$ is formulated by the product of group elements along a path between the base point and $v$.}
\label{fig:TorusExample}
\end{figure}

For the quantum double model on a torus, $g_{\alpha}$ are the holonomies along the two non-contractible cycles. See illustration in Fig. \ref{fig:TorusExample}. In particular in toric code model case, they corresponds to the operators $Z_1$ and $Z_2$.

The non-local transformation above defines a new basis of $\Hil^{B_p=1}$, $\{\ket{g_v;g_{\alpha}}\}$, which we call a \emph{holonomy basis}, as inspired by Ref.\cite{Wu1979}. In a holonomy basis, an operator $A_{v\neq v_0}$ acts as
\begin{equation}\label{eq:AvIng}
A_v\ket{g_1,\dots,g_v,\dots,g_{V-1};g_{\alpha}}
=\frac{1}{|G|}\sum_h\ket{g_1,\dots,hg_v,\dots,g_{V-1};g_{\alpha}},
\end{equation}
and $A_{v_0}$ acts as
\begin{equation}\label{eq:Av0}
A_{v_0}\ket{g_1,\dots,g_{V-1};g_{\alpha}}
=\frac{1}{|G|}\sum_h\ket{g_1\inv h,\dots,g_{V-1}\inv h;hg_{\alpha}\inv h}.
\end{equation}

\section{Fusion basis and the quasiparticle picture of entanglement}
\label{sec:qpPicture}
In this section, we propose a new picture of the entanglement between the two parts of a bipartite topologically-ordered system. In this picture, we can compute the entanglement entropy by counting the configurations of quasiparticles on the boundary between the two parts that divide the system.  We thus christen this new picture the \emph{quasiparticle picture}. We shall continue our study using the QD model.

Given a bipartite system, a state is entangled if it is not a tensor product of the states of the two subsystems. Entanglement entropy is a quantity that measures how much the two subsystems are entangled. 

For a given state $\psi$ in a bipartite system, the entanglement entropy is defined as
\be\label{eq:EE}
S_E=-\mathrm{tr} \rho_A\log \rho_A,
\ee 
where $\rho_A=\mathrm{tr}_B\ket{\psi}\bra{\psi}$ is the reduced density matrix. As such, the EE also measures one's ignorance of subsystem $B$, whose d.o.f. are traced out. One may also choose to trace out subsystem $A$ but $S_E$ is independence of such choices. In this paper, we focus on the EE in the ground states of the QD model.
\subsection{Entanglement entropy on a sphere in a holonomy basis}
\label{subsec:EEholoBasis}

To motivate the quasiparticle picture and understand how the EE is computed in the new picture, let us reexamine a well-studied case, namely, the EE of the QD model with a finite group $G$ on a sphere. In this case, the ground state is unique up to local unitary transformations, with trivial charge and zero flux everywhere on the sphere. This makes a holonomy basis handy for our computation of the EE. 

We virtually cut the sphere along an EB into a bipartite system. We would like to express the ground-state condition explicitly in terms of the charge d.o.f. on the EB. Doing this will manifest the contribution of quasiparticle fluctuations near the partition boundary to the EE and enable us to write down the reduced density matrix.

\begin{figure}[h]
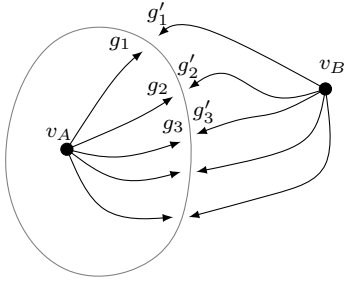

\centering
\GSSphere
\caption{Configuration of charge d.o.f. for the ground state of quantum double model on a sphere. There is a base point in each subsystem, with $v_A$ in $A$ and $v_B$ in $B$ respectively.}
\label{fig:GSSphere}
\end{figure}

As illustrated in Fig. \ref{fig:GSSphere}, we choose a base point $v_A$ ($v_B$) in subsystem $A$ ($B$). We assume that the EB consists of $L$ vertices. Since as far as the ground-state is concerned we can restrict the system to the subspace in which $B_p=1$ everywhere,  plus the sphere has no non-contractible cycles, the non-local transformation \eqref{eq:nonLocalTransform} would trade the original d.o.f. on the lattice edges for the new d.o.f. on the paths linking the base points to the vertices on the EB.
There should be two base points, one in subsystem $A$ and one in $B$. Otherwise, if there is just one base point chosen, say, in $A$, then there must be a path between any two neighbouring vertices on the EB, rendering all paths being closed. But $B_p=1$ everywhere, all such closed paths must be trivial, making no sense. 

Consequently, for each vertex on the EB, we have a charge d.o.f. $g_v$ in $A$ and $g'_v$ in $B$. A ground state should be gauge invariant at each $v$, and has zero holonomy everywhere. The latter condition reads
\begin{equation}\label{eq:zeroHolonomySphere}
\inv{g'_1}g_1=\inv{g'_2}g_2=\dots=\inv{g'_L}g_L.
\end{equation}

In terms of the new set of variables, the Hilbert space is spanned by the basis states $\ket{g_1,\dots,g_L;g'_1,\dots,g'_L}$. To obtain the ground state on the sphere, we can pick any basis state and act on it by all the vertex operators $A_{v\neq v_0}$ and $A_{v_0}$, which will project the basis state to a ground state. That is, pick a generic $\ket{g_1,\dots,g_L;g'_1,\dots,g'_L}$, then by Eqs. \eqref{eq:AvIng} and \eqref{eq:Av0}, we have
\be
\begin{aligned}
& A_{v_A} A_{v_B}\prod_{v=1}^L A_v \ket{g_1,\dots,g_L;g'_1,\dots,g'_L} \\
=& \sum_{h_1,\dots,h_L}\sum_{h_A,h_B}\ket{h_1 g_1 h_A^{-1},\dots,h_Lg_L h_A^{-1}; h_1 g'_1 h_B^{-1},\dots, h_L g'_L h_B^{-1}} \\
=& \sum_{\tilde g_1,\dots,\tilde g_L, h} \ket{\tilde g_1 h,\dots,\tilde g_L h;\tilde g_1,\dots,\tilde g_L}
,\end{aligned}
\ee
where $\tilde g_i := h_i g'_i h_B^{-1}$ and $h := h_B {g'}_1^{-1} g_1 h_A^{-1}$. The second equality above is due to the trivial holonomy condition after the action of the vertex operators, namely,
\[
h_i g_i h_A^{-1} = h_i g'_i h_B^{-1} (h_B {g'}_{i+1}^{-1} g_{i+1} h_A^{-1})=\tilde g_i (h_B {g'}_1^{-1} g_1 h_A^{-1})=\tilde g_i h,
\]
where the second equality is due to the 
trivial holonomy condition \eqref{eq:zeroHolonomySphere} before acting the vertex operators. By renaming $\tilde g_i$ back to $g_i$ again, we have a ground state 
\begin{align}
\ket{\Phi} &= \sum_{g_1,\dots,g_L,h}\ket{g_1h,\dots,g_Lh;g_1,\dots,g_L} \label{eq:GSstateSphereFull}\\
&= \sum_{g_v,h}\ket{g_v h;g_v} \label{eq:GSstateSphere},
\end{align}
where the second row follows the first row by an obvious suppression of indices. One can check that $\ket{\Phi}$ is indeed a ground state.

Since $g_v$ are the d.o.f. in region $B$, while $g_vh$ are the d.o.f. in region $A$, the form of $\ket{\Phi}$ is manifestly Schmidt-decomposed. Hence, the reduced density matrix $\rho_A$ is
\begin{equation}\label{eq:rhoSphere}
\rho_A=\frac{1}{|G|^2}\sum_{g_v,h,h'}\ketbra{g_vh}{g_vh'}.
\end{equation}
We check that $\rho_A$ is a projection operator
\begin{equation}\label{eq:rhoAProjectionSphere}
\begin{aligned}
\rho_A\rho_A=
&\frac{1}{|G|^4}\sum_{g_v,h,h'}\sum_{\bar g_v,\bar h,\bar h'} \ketbra{g_vh}{\bar g_v \bar h'}\braket{g_vh'}{\bar g_v\bar h'}\\
=&\frac{1}{|G|^4}\sum_{c}\sum_{g_v,h,\bar h'} \ketbra{g_vh}{g_v c^{-1}\bar h'}
\sum_{\bar g_v,\bar h,h'}
\delta_{\bar g_v^{-1}g_v,c}\delta_{\bar h' {h'}^{-1},c}\\
=&\rho_A.
\end{aligned}
\end{equation}
The normalization factor $\frac{1}{|G|^2}$ in Eq. \eqref{eq:rhoSphere} is chosen such that $\rho_A$ is exactly a projector. Note that since $\rho_A$ must represent a mixed state, if $\rho_A$ can be made a projector, it means that all probable pure-states are equally weighted in $\rho_A$. This leads to $S_E = -\mathrm{tr}(\rho_A\log\rho_A) =\log \mathrm{tr}\rho_A $. Now that
\begin{equation}\label{eq:traceRhoSphere}
\mathrm{W:=tr}\rho_A=\frac{1}{|G|^2}\sum_{g_v,h,h'} \delta_{h,h'}=|G|^{L-1},
\end{equation}
and the EE computes
as\begin{equation}\label{eq:SESphere}
S^{\text{sphere}}_E=\log W=(L-1)\log\ordG.
\end{equation}
This result agrees with the known result (See for example Ref.\cite{Chen2018}).
In the above we defined a quantity $W:=\mathrm{tr}\rho_A$. Clearly, Since all probable pure-states are equally weighted, $W$ is simply the number of probable pure-states. The questions are, what are these pure states and how we may interpret them? To answer the questions, it would be better to Fourier transform the holonomy bases into what we call the fusion bases.

%%%
\subsection{Fusion bases}
We have just derived the reduced density matrix and computed the corresponding EE of the QD model on a sphere. This is done in a holonomy basis proposed in Section \ref{subsec:holoBasis}. In the holonomy basis, the d.o.f. in either region $A$ or $B$ all live on the EB separating the two regions, and the d.o.f. in $A$ and those in $B$ group into entangled pairs, as seen in Eq. \eqref{eq:GSstateSphere}. This observation complies with the common sense that the entanglement between two regions actually occurs only near the EB between the two regions. It is also often said that the entanglement entropy measures the ignorance of an observer in region $A$ ($B$) about region $B$ ($A$). The computation in the holonomy basis not only manifests this common sense but also infers a new picture of entanglement, in which the ignorance can be made concrete in terms of the quasiparticle fluctuations on the EB. Imagine an observer in $A$. Tracing out region $B$ means that the observer in $A$ is completely refrained from knowing the existence of $B$, instead, the observer senses a boundary of $A$, which is the EB that now appears physical to the observer in $A$. Tracing out $B$ must leave certain residue on the EB. Recall that the entire system is in a ground state and that the total state \eqref{eq:GSstateSphere}, which is pure, is Schmidt decomposed in the holonomy basis as superposed entangled pairs. Since now the observer in $A$ cannot see region $B$ at all, he would likely see quasiparticle excitations on the EB. These excitations are paired up with those as appeared to an observer in region $B$ blinded from seeing region $A$, such that the entire system remains in a ground state. In the QD model, however, ($G$-charge type) quasiparticle excitations are better expressed in a basis not in terms of the group elements but in terms of the representations of the gauge group. As to be seen shortly, such a basis is a Fourier transform of the holonomy basis. Hence, what the observer in $A$ sees would not be a single type of quasiparticles but in fact a mixed state of all possible quasiparticle states on the EB. In other words, the observer sees quasiparticles fluctuating on the EB.     

In the following, we will apply the non-local transformation \eqref{eq:nonLocalTransform} and rewrite the reduced density matrix in terms of quasiparticle d.o.f..

We shall Fourier transform the holonomy basis in terms of group elements to one in terms of the irreducible representations of the gauge group. We\ label the charges by the irreducible representations. This will result in a more transparent picture of the quasiparticle fluctuations near the EB.

Define
\begin{equation}\label{eq:FourierTransform}
\ket{jmn}=\sqrt{\frac{d_j}{\ordG}}\sum_{g\in G}\overline{\rho^j_{mn}(g)}\ket{g},
\end{equation}
where $j$ labels the unitary irreducible representations of $G$, and $mn$ are matrix indices of the representations. The inverse transformation reads
\begin{equation}\label{eq:inverseFourierTransform}
\ket{g}=\sqrt{\frac{d_j}{\ordG}}\sum_{jmn}{\rho^j_{mn}(g)}\ket{jmn}.
\end{equation}
The above is the transformation on a state labeled by a single group element. The two transformations are unitary and inverse to each other, as can be proved by Peter-Weyl theorem in group representation theory. A state on the right hand side of the above equation can be regarded as a quasiparticle carrying the charge labeled by the irreducible representation. In the state \eqref{eq:GSstateSphere}, however, the holonomy basis consists of a chain of group elements for each of the two regions. Each such chain which would be transformed as in Eq. \eqref{eq:inverseFourierTransform}, resulting in a tensor product of states labeled by irreducible representations. The tensor product can be interpreted as the fusion of a chain of charge quasiparticles, leading naturally to a fusion basis $\ket{j_vm_v;\eta}$ of the states, depicted as
follows.\begin{equation}\label{eq:FusionBasisPic}
\FusionBasis,
\end{equation}
where $\eta$ labels the set of d.o.f. on the internal lines. These d.o.f. are intermediate quasiparticles in fusing of $L$ quasiparticles. See Appendix \ref{sec:FusionBasis} for the definition and the derivation of the fusion basis.\footnote{We note that such a basis was adopted in understanding the entanglement entropy in lattice gauge theories by one of us in Ref.\cite{HungWan2015}.}
In general, for a given chain of charge quasiparticles $j_v m_v$, there are multiple configurations $\eta$ of the intermediate quasiparticles. The number of such configurations is the dimension of the corresponding Hilbert space.

Note that the fusion basis \eqref{eq:FusionBasisPic} shows that the quasiparticles on the EB must all fuse to be a trivial charge. That is, the total quasiparticle charge of region $A$ including the EB is trivial. In fact, the reduced density matrix \eqref{eq:rhoSphere} is an identity matrix on all quasiparticle d.o.f. on the EB, combined with the projection $P_G$,
\begin{equation}\label{eq:rhoAPG}
\rho_A=P_G\left(\sum_{g_v}\ket{g_v}\bra{g_v}\right)P_G=P_G\mathds{1}P_G=P_G,
\end{equation}
where $P_G$ is the average of right multiplication on all $g_v$ by $h$. It enforces a $G$-symmetry. This global symmetry is a consequence of gauge invariance condition at the base point in $A$, which requires the total charge of all quasiparticles in $A$ to be trivial. The global $G$-symmetry can be viewed as the broken gauge symmetry due to the appearance of the EB. As a result, the nonzero eigenspace of $\rho_A$ are the space of all states of quasiparticles on EB that are invariant under this global $G$-symmetry. The $\eta$ is nothing but a set of good quantum numbers of this global symmetry.

In the fusion basis, Eq. \ref{eq:rhoSphere} becomes
\begin{equation}\label{eq:rhoj}
\rho_A=\sum_{j_vm_v,\eta}\ket{j_vm_v;\eta}\bra{j_vm_v;\eta},
\end{equation}
where $\eta,\eta'$ label the fusion basis with $j_vm_v$ fixed. Since the Fourier transform cannot prevent $\rho_A$ being a projector, the EE remains
\begin{equation}\label{eq:EEsphereFusionBasis}
S_E=\log{W},
\end{equation}
where $W$ is the dimension of the space spanned by the fusion basis. Directly counting the number of the basis vectors yields
\begin{equation}\label{eq:Wsphere}
\begin{aligned}
W=&\sum_{j_v}d_{j_1}\dots d_{j_L}\sum_{\eta_1,\dots,\eta_{L-1}}N_0^{j_1\eta_1}N_{\eta_1}^{j_2\eta_2}\dots N_{\eta_{L-1}}^{j_L0}\\
&=\sum_{\eta_1,\dots,\eta_{L-1}}d_0d_{\eta_1}^2\dots d_{\eta_{L-1}}^2 d_0\\
&=\ordG^{L-1},
\end{aligned}
\end{equation}
where use of $\sum_j N^{jk}_l d_j = d_k d_l$ is made in the second equality, and $j_1$ ($j_L$) is renamed to $\eta_1$ ($\eta_{L-1}$). See Appendix \ref{sec:FusionBasis} for more details. This result agrees with the previous result \eqref{eq:traceRhoSphere} obtained in the holonomy basis. This alternative derivation however has a deeper meaning, as we now elaborate on.

\subsection{Entanglement measures quasiparticle fluctuations}
\label{sec:QuasiparticleFluctuations}
The fusion-basis derivation in the previous subsection manifests a picture of entanglement in view of quasiparticle fluctuations. In this picture, we are able to answer the questions raised in the end of  Section \ref{subsec:EEholoBasis}.

In the language of gauge field theory, the ground state(s) of QD model satisfy two conditions---zero charge everywhere and zero flux everywhere. These two conditions determine the nonzero eigenspace of $\rho_A$, which can be spanned by the holonomy basis or the fusion basis developed in the previous subsections. In the following, we analyze the consequence of these two conditions on the charge d.o.f. $g_v$ and $g'_v$ on the EB .

Zero-flux condition requires the charge d.o.f. on both sides of the EB to be paired up to a global symmetry, as concluded in Eq. \eqref{eq:zeroHolonomySphere}. Here, we can imagine that the EB is fattened a bit to be  a narrow strip. In Fig. \ref{fig:GSSphere}, this requires $g_v=g'_vh$ for some constant $h$ at all  vertices $v$ on the EB. We shall consider the global symmetry on these charge d.o.f. later and let us ignore the constant $h$ at this moment. Locally, charges appear in pairs (and the pair goes across the fattened EB) at each vertex on the EB. We say that the charge quasiparticles in an EB-state satisfy the:
\begin{equation}\label{eq:boundaryMatchingConditiong}
\text{\it boundary matching condition}:\quad
g_v=g'_v,
\end{equation}
at all vertices $v$ on the EB.

Zero-charge condition has two consequences. First, at each vertex $v$ on the EB, charge pairs appear with equal weight. That is, locally, the ground state must read as $\sum_g\ket{g}_A\otimes\ket{g}_B$ because it is the only state invariant under gauge transformations at $v$. Hence, at each $v$ on the EB, the ground state locally reads as
\begin{equation}\label{eq:ChargePair}
\text{equal-weight charge pair}:\quad
\sum_{g_v}\ket{g_v}_A\otimes\ket{g_v}_B.
\end{equation}
The equal-weight charge pairs leads to a picture of quasiparticle fluctuations. A \emph{quasiparticle fluctuation} is a charge pair across the EB.
By the equal-weight condition \eqref{eq:ChargePair}, in the fusion basis these quasiparticles $j_vm_v$ (as well as their fusion multiplicity d.o.f. $\eta$) are paired across the EB. When the system is partitioned into two regions by the EB, however, local observers in a $A$ (in a mixed state $\rho_A$) can detect quasiparticles with equal probability on the EB. Our picture coincides with the quantum fluctuation in a vacuum state of a quantum field theory, which refers to a particle-antiparticle pair locally excited by an amount of energy. In our case, no actual energy is injected into the system, and a global observer (in a pure state $\ket{\Phi}$) cannot detect any quasiparticles anywhere because all quasiparticles on one side of the EB are confined with their counterparts on the other side. 

Second, the gauge-invariance condition at the base point of region $A$ imposes a global symmetry, implemented by the $P_G$ in Eq. \eqref{eq:rhoAPG}. All states $\ket{g_v}_A$ in $A$ ought to be invariant under the global right-multiplications over $G$. This global symmetry can be viewed as the broken gauge symmetry in $A$.
Since as appeared to a local observer in $A$ the EB is a PB, this gauge symmetry breaking is understood as due to the charge condensation on the EB\cite{Hung2013,Gu2014a}. 

To sum up, the ground state(s) of the QD model meet three conditions on the EB: the boundary matching condition, equal-weight local charge pairs, and a global symmetry $P_G$.

The advantage of the holonomy bases is that the d.o.f. are well allocated into two regions. By ``well'' we mean that all the three conditions on the ground state(s) are inherited by the EB states of region $A$ ($B$). In the following, we reexamine the three conditions from the eyes of the observers within $A$.

In the sphere case, the reduced density matrix $\rho_A$ is a projector, implying that the considered ground state $\ket{\Phi}$ admits a Schmidt decomposition:
\begin{equation}\label{eq:SchmidtSphere}
\ket{\Phi}=\sum_{i=1}^{W} \ket{\Psi_i}_A \otimes \ket{\Psi_i}_B.
\end{equation}
The states $\ket{\Psi_i}_A$ ($\ket{\Psi_i}_B$) are defined in region $A$ ($B$), rendering the total state $\ket{\Psi}$ explicitly an entangled state, as if region $A$ lies in a particular $\ket{\Psi_i}_A$, region $B$ has to be in the state $\ket{\Psi_i}_B$ indexed by the same $i$. Seen from the fusion-basis expression of the reduced density matrix \eqref{eq:rhoj}, $\ket{\Psi_i}_A = \ket{j_v m_v;\eta}$, where the index $i$ is a set of labels $(j_{v = 1,\dots,L}, m_{v = 1,\dots,L};\eta)$ specifying a particular fusion-basis state. In a state $\ket{\Psi_i}_A$, region $A$ has no quasiparticles everywhere except on the EB a particular configuration (e.g. \eqref{eq:FusionBasisPic}) of charge quasiparticles $j_v m_v$ with fixed internal d.o.f. labeled by $\eta$. Since the actual d.o.f. in region $A$ are the quasiparticles residing on the EB, we may abuse our terminology to call a $\ket{\Psi_i}_A$ an \emph{EB state} of region $A$. The elaboration above applies to region $B$ too.

Now let us adapt our analysis of the ground states to that of EB states $\ket{\Psi_i}_A$. The boundary matching condition \eqref{eq:boundaryMatchingConditiong}, together with the equal weight condition \eqref{eq:ChargePair} on chair pairs, implies that all possible charge quasiparticles will appear with equal probability on the EB in $A$. In the sphere case, this leads to that each of the $W$ basis-states $\ket{\Psi_i}_A$ is equally probable. 

The EB-states $\ket{\Psi_i}_A$ form a basis of all allowed states (in $A$) of the charge quasiparticles on EB. We summarize EB conditions on the EB states of region $A$ ($B$) as follows.
\begin{enumerate}
        \item They ensure that region $A$ ($B$) has no quasiparticles except on the EB.
        \item Locally, charge quasiparticles appear with equal probability everywhere on the EB.
        \item If there are non-contractible loops in region A (B), the holonomy along these loops is unit element of $G$.
        \item The total quasiparticle charge of either the entire region $A$ or $B$ is trivial, i.e., the quasiparticles in each allowed EB-state must fuse to be a trivial charge in either $A$ or $B$.
\end{enumerate}

Condition 3 is due to the special choice of the ground state which has trivial holonomy along these non-contractible loops. In general, if we choose other ground states $\ket{\Phi'}$, we should modify this condition such that any observable $\hat{O}$ along these non-contractible loops in region $A$ ($B$) should evaluates to the same value in $\ket{\Psi_i}_A$ ($\ket{\Psi_i}_B$) and in $\ket{\Phi'}$.

Condition 4 may look somehow odd at first glance that we require a global symmetry in $A$ and one in $B$ at the same time. We need both since they are the broken gauge symmetries in $A$ and in $B$. In the sphere case, the two global symmetries coincide with $P_G$ but in general they may differ as the spatial topology of the system becomes more complicated. We shall see such example in the next section and will discuss about this condition in more details in Sec. \ref{sec:GenericCase}.

Eq. \eqref{eq:SchmidtSphere} reads as a boundary matching condition upon gluing regions $A$ and $B$ along the EB, such that the total state $\ket{\Psi}$ is a ground state that is free of quasiparticles everywhere. 

\begin{figure}[ht!]
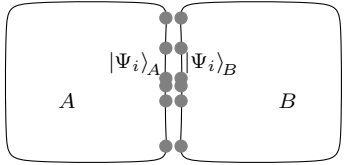

        \centering
        \SchmidtDecomp
        \caption{Schmidt decomposition of the ground state $\ket{\Phi}=\sum_i \ket{\Psi_i}_A \otimes \ket{\Psi_i}_B$. The entire system is divided into regions $A$ and $B$. The entangled EB-states $\ket{\Psi_i}_A$ ($\ket{\Psi_i}_B$) are defined for $A$ ($B$). Each EB-state specifies a configuration of charge quasiparticles on the EB. The EB-states of $A$ and those of $B$ must match, such that the total state of the entire system is a ground state.}
        \label{fig:SchmidtDecompositionSphere}
\end{figure}

Therefore, the EB-states of $A$ and those of $B$ are what are being entangled, as they are paired up. They are maximally entangled because all the EB-states are equally probable in the Schmidt decomposition. This physical picture is illustrated in Fig. \ref{fig:SchmidtDecompositionSphere}.

In terms of quasiparticles, the entanglement entropy entropy can be interpreted as a measurement of quasiparticle fluctuations. Quasiparticles on one side of the fattened EB are paired with their counterparts on the other side, and ``paired'' means ``entangled''. In fact, such pairing is the only contribution to the entanglement entropy, as we can see in the formula  $S_E=\log W$, which counts how many states of such quasiparticles are allowed on the EB. The fact that charge quasiparticles appear with equal probability on the EB is a consequence of a equilibrium state in region $A$ with \emph{maximal quasiparticle fluctuations} across the EB. We say an equilibrium state has ``maximal quasiparticle fluctuations'' if the energy threshold to excite any charge pairs on the EB becomes zero. (A ground state is an eigenvector of $A_v=1$ and reads locally as a superposition of equal-weight charge pairs as in Eq. \eqref{eq:ChargePair}. The partial trace $\text{tr}_BA_v$ becomes an identity operator, implying that quasiparticles on the EB can be excited without energy penalty as seen by the observers in $A$.) Hence we conclude:

\emph{Entangled spectrum corresponds to all possible states with quasiparticle fluctuations on the EB, and Entanglement entropy measures maximal quasiparticle fluctuations on the EB.}

In other words, \emph{entanglement entropy $S_E=\log W$counts the number of entangled EB-states of regions $A$ and $B$.} To compute $S_E$ amounts to the problem of state counting $W$ of the EB states satisfying the EB conditions.

The EE \eqref{eq:SESphere} consists of an leading area term $L\log |G|$ and a universal subleading term $-\log |G|$. The leading term results from the fact that there are $L$ vertices on the EB where quasiparticle pairs are allowed, and that there are $|G|$ types of quasiparticle pairs at each vertex (in either basis $\ket{g}$ or $\ket{jmn}$). The universal subleading term $-\log|G|$ results from the global symmetry imposed by $P_G$ in Eq. \eqref{eq:rhoAPG}, which says the quasiparticle fluctuations across the EB should preserve this global symmetry. The result agrees with our previous observation that the nonzero eigenspace of $\rho_A=P_G$ is the space of all states of quasiparticles on the EB that are invariant under the global $G$ symmetry.

\subsection{2D entanglement entropy = 1+1D thermal entropy}

The physical picture described above also demonstrates a correspondence between the EE and thermal entropy (TE). Suppose the total system is in a ground-state $\ket{\Phi}$. A local observer restricted in $A$ sees no existence of region $B$ but can measure the charge quasiparticles on the EB. The observer finds that there are $W$ equally probable quasiparticle configurations on the EB. Under the micro-canonical ensemble assumption, he computes the TE of the quasiparticle system on the EB as $\log W$. This result coincides with the EE we, as a global observer of the total system, obtained earlier. The local observer in region $A$ certainly does not know that the TE was a consequence of a hidden region $B$ but thought it was due to the quasiparticle fluctuation on the EB (a physical boundary as appeared to the observer. Note that the TE should be understood as one at infinite temperature because at $T=\infty$, all the quasiparticle configurations contribute to the partition function equally. We believe that the above correspondence in the sphere case can be generalized to a generic surface with non-trivial topology. If the total system is (may be in a mixed state due to topological ground state degeneracy) at zero temperature, and we compute $\rho_A$, then there exists certain effective Hamiltonian on the EB such that
\be
\rho_A = \lim_{T\rightarrow \infty}\e^{-H_{\mathrm{EB}}/T}.
\ee
The $H_{\text{EB}}$ is the effective Hamiltonian that causes the quantum fluctuations, and preserves the global symmetry broken from the gauge symmetries in $A$ and $B$. The infinite temperature agrees with our conclusion that the region $A$ is in equilibrium with maximal fluctuations on the EB.
This may be related to the program of modular flows (see Ref.\cite{Casini2011} and references therein).

\section{Topological orders with gapped physical boundaries}\label{sec:Cyl}
So far we have studied topological orders on closed two-spaces. Nevertheless, in reality, materials fabricated to bear topological orders usually have PBs. In particular, non-chiral topological orders, which can be studied by lattice Hamiltonian models, have gapped PBs between the system and the vacuum. It is therefore important to extend our study to the cases with physical boundaries. To this end, we consider the extended QD model with gapped boundaries\cite{Beigi2011,Bullivant2017}, which were developed to handle such cases. In this model, the bulk d.o.f. still take value in a finite gauge group $G$, while the d.o.f. on each PB take value in certain subgroup $K$ of $G$. It is said that the PB is characterized by $K$. Our holonomy bases and fusion bases naturally extend to such cases, and the picture of quasiparticle fluctuations still applies. To be specific, in what follows, we consider two representative cases, seen in Fig. \ref{fig:CylinerAA} and Fig. \ref{fig:CylinerAB}. The former
is a cylinder with region $A$ being a meridian strip, and the latter is a cylinder with region $A$ being a longitudinal strip. \subsection{Cylinder case I}
\begin{figure}[h]
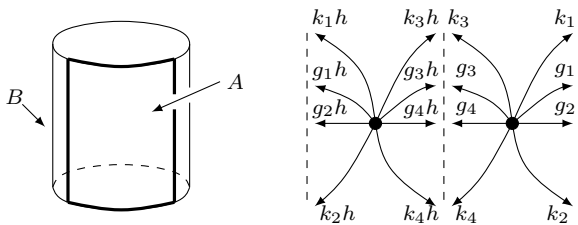

\centering
\CylinderAA
\quad
\CylinderAAState
\caption{A bipartite system on a cylinder. (a) Region $A$ is within the thick loop. (b) Configurations of group elements in the ground state chosen. The dash lines are the EBs. We take $k_1,k_3\in K_1$ and $k_2,k_4\in K_2$. We place the $k$'s and $g$'s in one region and their duplicates in the other, as a consequence of trivial holonomy condition.}
\label{fig:CylinerAA}
\end{figure}

Consider the extended QD model on the cylinder in Fig. \ref{fig:CylinerAA}(a). The model is defined by a QD Hamiltonian in the bulk with gauge group $G$, and by a boundary Hamiltonian\cite{Bullivant2017} on PBs specified by two subgroups $K_1$ and $K_2$ of $G$. The subgroup $K_1$ ($K_2$) dictates the gapped PB condition on the top (bottom) PB of the cylinder. Topological GSD occurs in general on a cylinder. As to be specified below, we will focus on only one particular ground state, which bears no non-contractible anyon loops.

Similar to the case on a sphere, we choose a base point $v_A$ ($v_B$) in $A$ ($B$). Since trivial holonomy is assumed everywhere, we only need to consider the d.o.f. on the EB, which is a disjointed union of two vertical line segments, each with two end vertices respectively on the top and bottom PBs. See Fig. \ref{fig:CylinerAA}(b) for an illustration.

By a derivation similar to that for the ground state on a sphere, we can find one particular ground state (out of many others), which does not have any non-contractible loops, as
\begin{equation}\label{eq:GSCylinder}
\ket{\Phi}_{\text{cyl}}=\sum_{g_v,h,h'\in G,k_1,k_3\in K_1,k_2,k_4\in K_2}\ket{k_uh,g_vh;k_uh',g_vh'},
\end{equation}
where $v$ labels the bulk vertices on the EB, and $u=1,3$ ($u=2,4$) label the vertices on the intersections of the upper (lower) PB and EBs. One can verify the $K_1,K_2$-gauge invariance at the PB vertices $u$ and $G$-gauge invariance at $v$. Hence, $\ket{\Phi}_{\text{cyl}}$ is indeed a ground state.
Explicitly, the state is obtained by applying average (bulk and boundary) gauge transformations everywhere on the state with $k_u,g_v$ set to be $1$.

As in the sphere case, the ground state $\ket{\Phi}_{\text{cyl}}$ is manifestly Schmidt decomposed; hence, the reduced density matrix is readily
\begin{equation}\label{eq:rhoCylinderI}
\rho_A=\sum_{g_v,h,h'\in G,k_1,k_3\in K_1,k_2,k_4\in K_2}\ketbra{k_uh,g_vh}{k_uh',g_vh'},
\end{equation}
where $K=K_1\cap K_2$. 
Similar as in the sphere case, one can show that $\rho_A$ is a projector multiplied by a normalization factor. Direct calculation leads to a normalized reduced density matrix
\begin{equation}\label{eq:DerivenormalizedrhoCylinderI}
\bar \rho_A=\frac{1}{|G||K|}\rho_A,
\end{equation}
satisfying
\begin{equation}\label{eq:normalizedrhoCylinderI}
\bar{\rho}_A\bar{\rho}_A=\bar{\rho}_A,
\quad
\mathrm{tr}\bar{\rho}_A=W=\frac{|G|^L|K_1|^2|K_2|^2}{|K|},
\end{equation}
where $L$ is the total number of bulk vertices on the EB. The EE is
\begin{equation}\label{eq:EECylinderI}
S_E=\log W=\log\frac{|G|^L|K_1|^2|K_2|^2}{|K|}.
\end{equation}
See Appendix \ref{sec:rhoCylinderI} for the detailed evaluation. Despite in this case the holonomy basis \eqref{eq:GSCylinder} has already made the EE computation simple, we would still like to write down the fusion basis as follows to manifest the picture of quasiparticle fluctuations.

\subsubsection{The fusion basis}

By examining Eq. \eqref{eq:rhoCylinderI} and \eqref{eq:normalizedrhoCylinderI}, we can extract a picture of quasiparticle fluctuations. As illustrated in Fig. \ref{fig:CylinerAA}, The EB intersects PB at the four vertices labeled by $u$. The reduced density matrix consists of two projectors. One projector is
\begin{equation}\label{eq:P0}
P_0=\sum_{k_u,g_v}\ket{k_u,g_v}\bra{k_u,g_v},
\end{equation}
which projects onto the subspace where we assign $k_1,k_3\in K_1$ on the upper PB and $k_2,k_4\in K_2$ on the lower PB. At other vertices on the EB, we assign group elements $g_v\in G$. The other projector is an average of global right-multiplications:
\begin{equation}\label{eq:PG}
P_G=\sum_{g_u,g_v,h\in G}\ket{g_uh,g_vh}\bra{g_u,g_v}
\end{equation}

The reduced density matrix is a combination:
\begin{equation}\label{eq:rhoACylinderIApp}
\rho_A=P_GP_0P_G.
\end{equation}
This formula is similar to that in the sphere case, where the identity matrix in Eq. \eqref{eq:rhoAPG} is replaced by the projector $P_0$. As suggested by the projector $P_0$, a good fusion basis should be constructed in terms of the irreducible representations of $K_1,K_2$ at the four vertices $u$ and the irreducible representations of $G$ at the other vertices $v$. The projector $P_G$ enforces a global right-multiplication symmetry, which leads to a fusion tree structure similar to Eq. \eqref{eq:FusionBasisPic}. To be clear, we draw the fusion basis as
\begin{equation}\label{eq:fusionBasisCylinderI}
\FusionBasisCylinder
\end{equation}
where $p_1,p_3$ ($p_2,p_4$) label the irreducible representations of $K_1$ ($K_2$), and the $j$'s are those of $G$. The $x$'s and $m$'s are the internal indices of the corresponding representations. Residing on the two big dots are the unitary transformations $U^j_{p,\alpha}$, defined in Eq. \eqref{eq:UU}, that decompose $G$ representations to $K_{1,2}$ representations. On internal lines and trivalent vertices are the invariant tensor basis as defined in Eq. \eqref{eq:etaeta}, for $K_1$, $K_2$, and $G$ respectively.

It is more complicated to solve the eigen-problem of $\rho_A$ in such fusion basis than in the holonomy basis. One complexity is that such fusion basis is not orthogonal with respect to the d.o.f. at the two big dots (in terms of $U^j_{p,\alpha}$). (These dots are $K_{1,2}$-morphisms but not $G$-morphisms, and hence are not invariant under $P_G$.) We will not detail the computation of EE in such fusion basis.

The entanglement entropy in Eq. \eqref{eq:EECylinderI} consists of two terms,
\begin{equation}\label{eq:EECylinderIqp}
S_E=S_0+S_1,
\end{equation}
with the leading area term
\begin{equation}\label{eq:EECylinderIleading}
S_0=L \log|G|+2\log|K_1|+2\log|K_2|,
\end{equation}
where $L$ is the length of the EB in the bulk, i.e., the number of bulk vertices on the EB. The latter two terms reflects the contribution of quantum fluctuations at the four vertices at the intersections of the EBs and the PBs. Each such vertex on the top (bottom) PB allows $|K_1|$ ($|K_2|$) types of quasiparticles.
The subleading term is
\begin{equation}\label{eq:EECylinderIsubleading}
S_1=-\log|K|,
\end{equation}
as a consequence of the global $G$-symmetry on all quasiparticles on the EB (including the four intersection points).

\subsection{Cylinder case II}

\begin{figure}[ht!]
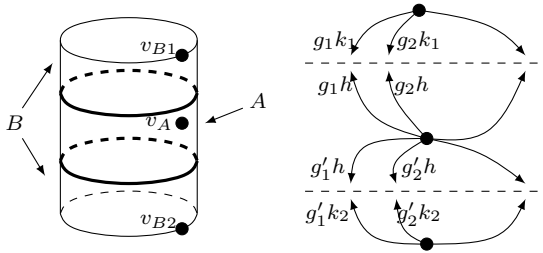

\centering
\CylinderAB
\CylinderABState
\caption{A bipartite system on a cylinder. (a) Region $A$ is bounded by the thick line. (b) Configurations of group elements for a ground state. The dash lines are the two EBs. The group elements $k_1\in K_1$, $k_2\in K_2$, and  $g_v,g'_v,h\in G$. We place the $k$'s and $g$'s in one region and their duplicates in the other, as a consequence of trivial holonomy condition.}
\label{fig:CylinerAB}
\end{figure}

We consider another bipartite system on a cylinder as in Fig. \ref{fig:CylinerAB}(a).
The extended QD model is the same as in the cylinder case I, where the bulk gauge group is $G$, and the top and bottom PB conditions are characterized by subgroups $K_1$ and $K_2$ of $G$. Again we will focus on only one particular ground state to be specified below.

Since region $B$ now has two disjoint components, we will choose the base points in a way slightly different from that in the previous case. We pick two base points in region $B$, with $v_{B1}$ on the top PB and $v_{B2}$ on the bottom PB. See Fig. \ref{fig:CylinerAB}(a). The d.o.f. to specify a ground state are shown in Fig. \ref{fig:CylinerAB}(b).

The ground state of our concern takes the form
\begin{equation}\label{eq:GSCylinderII}
\ket{\Phi}=\sum_{k_1\in K_1,k_2\in K_2,g_v,g'_v,h\in G}\ket{g_vh,g'_vh;g_vk_1,g'_vk_2}.
\end{equation}
The reduced density matrix is
\begin{equation}\label{eq:partialtraceCylinderII}
\begin{aligned}
\rho_A=&\mathrm{tr}_B\ket{\Phi}\bra{\Phi}\\
=&\sum_{g_v,g'_v,h,\bar h}\sum_{c_1\in K_1,c_2\in K_2}
\ketbra{g_vh,g'_vh}{g_vc_1h',g'_v c_2\bar h}\\
&\sum_{\bar g_v,\bar g'_v,k_1k_2\bar k_1\bar k_2}
\delta_{g_v^{-1}\bar g_v,c_1}\delta_{k_1{\bar k_1}^{-1},c_1}
\delta_{{g'_v}^{-1}\bar g'_v,c_2}\delta_{k_2{\bar k_2}^{-1},c_2}.
\end{aligned}
\end{equation}
which evaluates as (up to a normalization factor)
\begin{equation}\label{eq:rhoCylinderII}
\rho_A=\sum_{k_1\in K_1,k_2\in K_2,g_v,g'_v,h,h'\in G}\ketbra{g_vh,g'_vh}{g_vk_1h',g'_vk_2h'}
\end{equation}

In particular when $G$ is Abelian, $\rho_A$ can be normalized to be a projector
\begin{equation}\label{eq:projectionCylinderII}
\bar\rho_A=\frac{1}{|G|^2|K_1||K_2|}\rho_A,
\quad
\bar\rho_A\bar\rho_A=\bar\rho_A,
\end{equation}
with trace being
\begin{equation}\label{eq:traceCylinderII}
W:=\mathrm{tr}\bar\rho_A=\frac{|G|^{L-1}|K|}{|K_1||K_2|},
\end{equation}
where $K=K_1\cap K_2$. See Appendix \ref{sec:rhoCylinderII} for the detailed computation. The EE in the Abelian case is thus
\begin{equation}\label{eq:EECylinderIIAbelian}
S_E=\log W=\log\frac{|G|^{L-1}|K|}{|K_1||K_2|},
\end{equation}
which agrees with the result in Ref.\cite{Chen2018}.

In the non-Abelian case, we need some more tricks to obtain a closed-form formula of the EE. We leave the analytic computation in the fusion basis to the next subsection.

\subsubsection{The fusion basis}\label{subsubsec:CylIIfusion}

We see that $\rho_A$ is a projector. We also see that $\rho_A$ is gauge invariant at all of the base points chosen, in kets and bras respectively. These facts imply all four conditions on the EB-states are fulfilled. In this subsection, we analyze the entanglement in the picture of quasiparticle fluctuations using the fusion basis.

Let
\begin{equation}\label{eq:defineL}
R_{g_1,g_2}=\sum_{g_v,g'_v}\ket{g_vg_1,g'_vg_2}\bra{g_v,g'_v}
\end{equation}
be the right-multiplication acting on $g_v$ and $g'_v$ by $g_1$ and $g_2$. Define two projectors
\begin{equation}\label{eq:definePG}
P_G=\frac{1}{|G|}\sum_{g\in G}R_{g,g}
\end{equation}
\begin{equation}\label{eq:definePKK}
P_{K_1}=\frac{1}{|K_1|}\sum_{k_1\in K_1}R_{k_1,1},
P_{K_2}=\frac{1}{|K_2|}\sum_{k_2\in K_2}R_{1,k_2}.
\end{equation}

Then the reduced density matrix can be written as
\begin{equation}\label{eq:rhoAinP}
\rho_A=P_GP_{K_1K_2}P_G,
\end{equation}
where $P_{K_1K_2}=P_{K_1}P_{K_2}$.
If we traced out region $A$, we would get
\begin{equation}\label{eq:rhoBinP}
\rho_B=P_{K_1K_2}P_GP_{K_1K_2}.
\end{equation}

The projectors impose the global constraints in the bulk and on the PB respectively. The $P_G$ projects onto the states with zero charge in bulk, while $P_{K_1}$ and $P_{K_2}$ projects onto the states with zero boundary $k_{1,2}$ charge at each boundary component.

We now try to find the eigenvectors of $\rho_A$. First consider the effect of $P_G$. Let us take a fusion basis, as obtained in Eq. \eqref{eq:FusionBasisPic}:
\begin{equation}\label{eq:CylinderIIfusionbasisApp}
\FusionBasisCylinderII
\end{equation}
where we divide the labels  $j_1m_1,\dots,j_Lm_L$ into two sets, with the d.o.f. on the upper (lower) EB all on the left (right) hand side of $\eta_0$. Between these two parts there is one internal line labeled by $\eta_0$, an irreducible representation of $G$. The combined projector $P_{K_1K_2}$ is a projection on $\eta_0$ and contribute a factor
\begin{equation}\label{eq:factorCylinderII}
N_{\eta_0}=\frac{1}{|K_1||K_2|}\frac{1}{\dim_{\eta_0}}\sum_{k_1k_2}\chi_{\eta_0}(k_1k_2),
\end{equation}
Which satisfies
\begin{equation}\label{eq:NetaProperty}
\sum_{\eta_0}d_{\eta_0}^2N_{\eta_0}=|G|/|K_1K_2|.
\end{equation}

The eigenvalues are $N_{\eta_0}$, whose degeneracies are calculated in a way similar  to that in Eq. \eqref{eq:Wsphere},
\begin{equation}\label{eq:NetaDegeneracy}
\begin{aligned}
W_{\eta_0}=&\sum_{j_v}d_{j_1}\dots d_{j_L}\sum_{\eta_1,\dots,\hat{\eta_0},\dots,\eta_{L-1}}N_0^{j_1\eta_1}N_{\eta_1}^{j_2\eta_2}\dots N_{\eta_{L-1}}^{j_L0}\\
&=\sum_{\eta_1,\dots,\hat{\eta_0},\dots,\eta_{L-1}}d_0d_{\eta_1}^2\dots d_{\eta_{L-1}}^2 d_0\\
&=\ordG^{L-2}d_{\eta_0}^2,
\end{aligned}
\end{equation}
where $\hat{\eta_0}$ in the sum means that $\eta_0$ is a fixed label and is not summed.

The trace evaluates as
\begin{equation}\label{eq:trWeta}
W=\mathrm{tr}(\rho_A)=\sum_{\eta_0}W_{\eta_0}N{\eta_0}=\frac{|G|^{L-1}}{|K_1K_2|}.
\end{equation}
The normalized eigenvalues are then $\lambda_{\eta_0}=N_{\eta_0}/W$ with degeneracy $W_{\eta_0}$, and hence the EE computes as
 \begin{equation}\label{eq:EEcylAB}
\begin{aligned}
S_E=&-\sum_{\eta_0}W_{\eta_0}\lambda_{\eta_0}\log \lambda_{\eta_0}
\\
=&-\sum_{\eta_0}W_{\eta_0}\frac{N_{\eta_0}}{W}\log \frac{N_{\eta_0}}{W}\\
=&\log W-\sum_{\eta_0}\frac{d_{\eta_0}^2|K_1K_2|}{|G|}N_{\eta_0}\log N_{\eta_0}.
\end{aligned}
\end{equation}

The EE again contains two terms, $S_E=S_0+S_1$, with the leading area term $S_0=L \log|G|$ and a subleading term
\begin{equation}\label{eq:EESubleadingCylinderII}
S_1=-\log|G|-\log|K_1K_2|-\sum_{\eta_0}\frac{d_{\eta_0}^2|K_1K_2|}{|G|}N_{\eta_0}\log N_{\eta_0},
\end{equation}
as a consequence of three mutually un-commuting symmetries: $P_G$ acting on all quasiparticles and $P_{K_1}$ ($P_{K_2}$) acting on the quasiparticles on the upper (lower) EB.

This closed-form formula is a novel result, which has not been obtained in other works on the EE of the QD model. In a recent work\cite{Chen2018}, a numeric result of the EE in this case for $G=S_3$, $K_1=\{1,4\}$, and $K_2=\{1\}$ was given. Let us now verify as an example that our formula \eqref{eq:GSCylinder} yields the same numeric result. In this example, we have
\begin{equation}\label{eq:Netaexample}
N_1=1,N_2=0,N_{3}=1/2.
\end{equation}
See Appendix \ref{sec:example} for $S_3$ and its representations.
Eq. \eqref{eq:GSCylinder} quickly leads to
\begin{equation}\label{eq:Wetaexample}
S_E=\log |G|^{L-1}-\frac{1}{3}\log 2,
\end{equation}
which is precisely the result in Ref.\cite{Chen2018}.

When $G$ is Abelian, $N_\eta$ is either $0$ or $1$ (as can be directly verified from the definition in Eq. \eqref{eq:factorCylinderII}). Hence the last term in Eq. \eqref{eq:EESubleadingCylinderII} vanishes, recovering the result \eqref{eq:EECylinderIIAbelian}.

\subsection{Generic cases}\label{sec:GenericCase}

\begin{figure}[!ht]
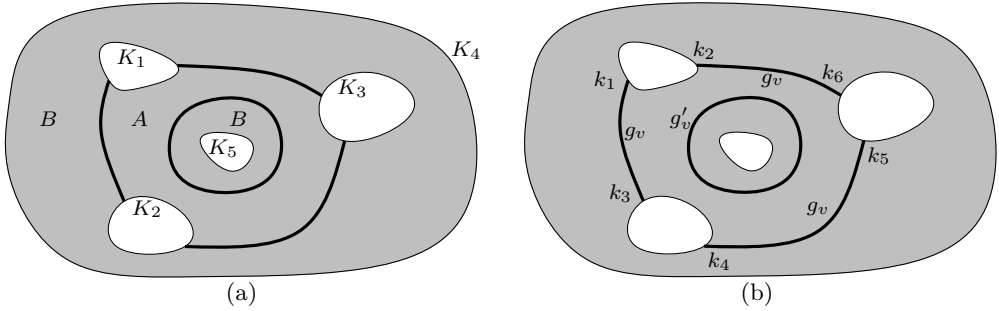

        \centering
        \subfigure[]{\GenericSurface}\label{fig:GenericSurfaceA}
        \subfigure[]{\GenericSurfaceConfig}\label{fig:GenericSurfaceB}
        \caption{(a) An open surface with boundaries. The thick lines are the EBs that separate the system into two regions $A$ and $B$. The five PBs are specified by the subgroups $K_1,K_2,K_3,K_4$, and $K_5$ of $G$. (b) The subspaces with group elements $k_1,k_2\in K_1$, $k_3,k_4\in K_2$, and $k_5,k_6\in K_3$, and $G$ elements $g_v$ and $g'_v$ assigned to the EBs.}
        \label{fig:GenericSurface}
\end{figure}

We have discussed the picture of quasiparticle fluctuations in two representative cases on a cylinder; however, this picture is valid on more generic surfaces. We will summarize some typical features of the previous analysis and then extend them to a generic surface. For example, consider an open surface with PBs as illustrated in Fig. \ref{fig:GenericSurfaceA}. We do not consider any genus in the bulk because we always choose a particular ground state such that the global (topological) d.o.f. will not affect our computation. Specifically, we choose a ground state in which the holonomy along all non-contractible loops are the unit element of $G$, which makes any genus invisible in the holonomy bases.

Let us try to construct the reduced density matrices on  generic surfaces. We can see that the reduced density matrices \eqref{eq:rhoAPG}, \eqref{eq:rhoACylinderIApp}, and \eqref{eq:rhoAinP} all are a combination of three types of projectors. The first type includes projectors due to the boundary conditions on the PBs, such as the $P_0$ in Eq. \eqref{eq:P0}, projecting all group elements at the intersection of PBs and EBs into the characterizing subgroups for PBs. (This type of projectors also depends on the choice of the ground state but we have already restricted to ground states without anyon loops throughout the paper.) The other two types are due to the gauge symmetries in regions $A$ and $B$ respectively, such as the $P_G$ and $P_{K_1K_2}$ in Eq. \eqref{eq:rhoAinP}, where $G$ is the gauge group in $A$, and $K_1$ ($K_2$) is the gauge group of the top (bottom) PBs in $B$. 

These observations comply with the four conditions on the EB in Sec \ref{sec:QuasiparticleFluctuations}. Now we may formulate these conditions mathematically by writing down a generic form of the reduced density matrix.

Keep Fig. \ref{fig:GenericSurface} in mind as an example, we itemize the rules to write down the reduced density matrix:
\begin{enumerate}
        \item The first projector projects onto a subspace with only $K_1$, $K_2$ and $K_3$ d.o.f. at all intersections between EBs and PBs, and assigns $G$-elements elsewhere. In this example, it is written as
        \begin{equation}\label{eq:PGenericAA}
        P_0=\sum_{k_1k_2\in K_1,k_3k_4\in K_2,k_5k_6\in K_3}\ket{k_u,g_v,g'_v}\bra{k_u,g_v,g'_v},
        \end{equation}
        where $u=1,2,\dots,6$.
        \item Suppose region $B$ has a number of disconnected components. For any disconnected component that contains a PB characterized by $K$, we write down a projector as the average over $K$ of the right-multiplications acting on the EBs that are a part of the boundary of this disconnected component. For a disconnected component that does not contain any PB, we write down the projector $P_G$. (For simplicity, we do not consider the case with multiple PBs contained within each component of $B$) In the current example, $B$ has two disconnected components, and the projectors are
        \begin{equation}\label{eq:PGnericBB}
        P_{K_4}=\sum_{l\in K_4}\ket{lg_u,lg_v,g'_v}\bra{g_u,g_v,g'_v},
        \end{equation}
        and
        \begin{equation}\label{eq:PGnericCC}
        P_{K_5}=\sum_{l\in K_5}\ket{g_u,g_v,lg'_v}\bra{g_u,g_v,g'_v}.
        \end{equation}
        Since these projectors are due to the global symmetry in region $B$, we combine them as the projector
        \begin{equation}\label{eq:PGenericPB}
        P_B=P_{K_4}P_{K_5}.
        \end{equation}
        \item Repeat the second step also for all disconnected components of $A$. In the current example, we have
        \begin{equation}\label{eq:PGnericDD}
        P_A=P_G=\sum_{h\in G}\ket{hg_u,hg_v,hg'_v}\bra{g_u,g_v,g'_v}.
        \end{equation}
        \item We finally combine all the projectors and obtain
        \begin{equation}\label{eq:rhoAGneric}
        \rho_A=P_AP_BP_0P_BP_A.
        \end{equation}
\end{enumerate}

This form of $\rho_A$ is a consequence of the four conditions on EB-states.  Conditions 1, 2, and 3 allow equal-probable configurations of charge quasiparticles on the EB. Hence, we arrive at the $P_0$ in Eq. \eqref{fig:GenericSurfaceA}. Under condition 4, the two global symmetry constraints in $A$ and $B$ implies two projection operators $P_A$ in Eq. \eqref{eq:PGnericDD} and $P_B$ in Eq. \eqref{eq:PGenericPB}. Combining all four conditions we arrive at the formula for $\rho_A$ in Eq. \eqref{eq:rhoAGneric}.

\begin{figure}[!ht]
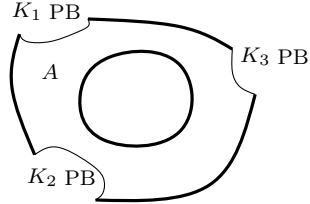

        \centering
        {\GenericSurfaceAA}
        \caption{Observers in region $A$ ($B$ is traced out) think all PBs and EBs are physical boundaries.}
        \label{fig:GenericSurfaceAA}
\end{figure}

Before explaining how we combine the projectors $P_0$, $P_A$ and $P_B$ in such a way to get $\rho_A$, let us figure out how the observers in $A$ would view their world when $B$ is traced out. They cannot distinguish a PB from an EB but think all the PBs and EBs are physical boundaries of their world. See Fig. \ref{fig:GenericSurfaceAA}. By Condition 4 on EB-states, we have a global symmetry acting on the EB d.o.f. due the broken gauge symmetry in $B$. The resulting global symmetry is implemented by $P_B$. Observers in $A$ cannot understand why they can observe such a global symmetry on a boundary that was an EB because they are blinded from $B$. In this way, the projector $P_B$ encodes how the topology and PBs of $B$ may affect the EB-states observed in $A$.

There is another global symmetry acting on the EBs. Combining the already existing PB conditions (characterized by $K_1$, $K_2$, and $K_3$) and the effective boundary conditions (expressed as $P_B$) on the EBs, the gauge symmetry in $A$ also breaks into a global symmetry, now imposed by $P_A$. The above interpretation of $P_A$ and $P_B$ enables us to write down the combination of the projections in the order in Eq. \eqref{eq:rhoAGneric}.

We may choose to trace out region $A$ instead and get
\begin{equation}\label{eq:rhoBGeneric}
\rho_B=P_BP_AP_0P_AP_B,
\end{equation}
in which the order of projectors accords with the above interpretation of the projectors.

Since $P_A$ and $P_B$ do not commute, $\rho_A\neq\rho_B$. The nonzero eigenspaces of $\rho_A$ and $\rho_B$ are not identical. This is reasonable because the regions $A$ and $B$ have different spatial topologies and different PB conditions. The local observers in $A$ and $B$ can distinguish such differences in the EB-states. For example, local observers will find the constraint \eqref{eq:factorCylinderII} on the fusion basis \eqref{eq:CylinderIIfusionbasisApp} in Cylinder case II, and find the constraint on the dotted part in fusion basis \eqref{eq:fusionBasisCylinderI}. Nevertheless, the EE computed from $\rho_A$ and that from $\rho_B$ out to be equal.

\subsubsection{The disk case}
\begin{figure}[!ht]
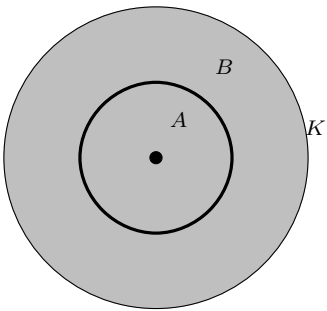

        \centering
        {\Disk}
        \caption{A bipartite system on a disk. Region $A$ is bounded by the thick line. The PB condition is characterized by a subgroup $K$ of $G$.}
        \label{fig:Disk}
\end{figure}

As a simple example, we apply our generic formula \eqref{eq:rhoAGneric} to the disk case. We consider the extended QD model a disk as in Fig. \ref{fig:Disk}(a).
The bulk gauge group is $G$, and the PB is specified by a subgroup $K$ of $G$. The system is bipartite by an EB (solid circle in the figure). The ground state on the disk is unique, as to be specified below.

The d.o.f. would be $g_v$ for all $v$ on the EB. The global symmetries would be $P_A=P_G$ being the average of right-multiplications over $G$ on $g_v$ for all $v$, and $P_B=P_K$  over $K$. Hence the reduced density matrix $\rho_A$ is
\begin{equation}\label{eq:PPPDisk}
\rho_A=P_GP_K\mathbf{1}P_KP_G=P_G.
\end{equation}
In the second equality we have used $P_GP_K=P_KP_G=P_G$. Hence the entanglement spectrum and entropy are the same as those in the sphere case.

\section{Discussions}\label{sec:disc}

In this paper, we always choose a ground state in which there are no non-contractible anyon loops. In the sphere case, certainly there are no such loops anyway but on a cylinder, there may be because a topological order on a cylinder may have ground state degeneracy\cite{HungWan2014,Bullivant2017,Hu2017a}. In general, we can choose an arbitrary ground state, with or without non-contractible anyon loops. In such cases, while the analysis and the main result will remain unaffected, we would need to slightly extends the four conditions on the entangled EB-states. Moreover, one may also consider topological orders with excited anyons, i.e., not in ground states. This would not affect the physics of EE, although special cares may be needed for defining the entangled EB-states. We shall leave such cases for future work.

In this paper, each EB accommodates only the quasiparticles that are pure charges. The reason is that each EB is composed of consecutive edges of the lattice; hence, the boundary condition is specified by $G$ itself. This boundary condition corresponds to the smooth boundary condition on the PBs in the extended QD model\cite{Beigi2011,HungWan2015,Bullivant2017}. Under the smooth boundary condition, the gapped boundary is a consequence of condensing all the flux quasiparticles on the PB, such that the gapped quasiparticle excitations on the PB are charges only. The correspondence between our EB and flux condensation on the PB of the extended QD model is more transparent to Alice, the local observer in region $A$, to whom the EB is effectively a PB. There are other choices of EBs, on which the quasiparticles are either pure fluxes, dyons, or of mixed types. According to Ref.\cite{HungWan2015}, the TEE of topological orders  is independent of the choice of EB. We thus do not consider other choices of EBs here. Moreover, in Ref.\cite{Hung2019}, the authors computed the EE between two different topological orders separating by a gapped domain wall by treating the domain as the EB. Since the gapped domain wall is a result of anyon condensation on the wall, the correspondence between our EBs and anyon condensation is natural.

Although our study was done for the (extended) QD model only, we believe our results can be extended to other models of topological phases, such as the twisted quantum double model\cite{Hu2012a,Bullivant2017}, the Levin-Wen model\cite{Levin2004,Lin2014,Hu2017,Hu2017a,Hu2018} in two dimensions, and the twisted gauge theory model in three dimensions\cite{Wan2014,Wang2018a}. In such extensions, however, the dimension of group representations should be extended to the quantum dimensions of the anyons. The latter is more general measure in the modular tensor category theories---the general algebraic theory of quasiparticles. We shall leave the generalization for future work.

\appendix

\section{Fusion basis in the sphere case}\label{sec:FusionBasis}
We have used the fusion bases to represent the EB-states. Here we provide a concrete definition of the fusion bases. We define a fusion basis by
\begin{equation}\label{eq:FusionBasisSphere}
\begin{aligned}
&\ket{j_1m_1,j_2m_2,\dots,j_Lm_L;\eta}\\
:=&\sum_{n_1,n_2,\dots,n_L}T^\eta_{j_1n_1,j_2n_2,\dots,j_Ln_L}\ket{j_1m_1n_1,j_2m_2n_2,\dots,j_Lm_Ln_L},
\end{aligned}
\end{equation}
where $T^\eta_{j_1m_1,j_2m_2,\dots,j_Ln_L}$ is a tensor transforming under irreducible representations of $G$:
\begin{equation}\label{eq:tensorSphere}
\sum_{n'_i}\rho^{j_i}_{n_in'_i}T^\eta_{j_1n_1,j_2n_2,\dots,j_in'_i,\dots,j_Ln_L}=T^\eta_{j_1n_1,j_2n_2,\dots,j_in_i,\dots,j_Ln_L}.
\end{equation}
This generalizes the Wigner-Eckart tensors with arbitrary number of pairs $jn$.
The index $\eta$ labels a normal basis of such tensors such that the contraction of tensors satisfy
\begin{equation}\label{eq:etaeta}
T^{\eta}\cdot\overline{T^{\eta'}}=\delta_{\eta \eta'},
\end{equation}
and
\begin{equation}\label{eq:sumeta}
\sum_{\eta}T^\eta_{j_1m_1,j_2m_2,\dots,j_Lm_L}\overline{T^\eta_{j_1n_1,j_2n_2,\dots,j_Ln_L}}
=\delta_{m_1n_1}\delta_{m_2n_2}\dots\delta_{m_Ln_L}.
\end{equation}
From the group representation theory, such $T^{\eta}$ can be constructed as contractions of $L-3$ $3j$-symbols (for $L>3$)
\begin{equation}\label{eq:TetaThreeJsymbols}
T^{\eta_1,\dots,\eta_{L-3}}_{j_1m_1,\dots,j_Lm_L}=
\sum_{x_1,\dots,x_{L-3}}
C^{j_1j_2\eta_1}_{m_1m_2x_1}C^{\eta_1^*j_3\eta_2}_{x_1m_3x_2}\dots C^{\eta_{L-3}^*j_{L-1}j_L}_{x_{L-3}m_{L-1}m_L},
\end{equation}
where we label $\eta$ by a set of irreducible representations $\eta_1,\dots,\eta_{L-3}$ of $G$.

By the above construction, we simply present the fusion basis \eqref{eq:FusionBasisSphere} graphically as
\begin{equation}\label{eq:FusionBasisPicAppendix}
\FusionBasis
\end{equation}

\section{Reduced density matrix in cylinder case I}
\label{sec:rhoCylinderI}
We show that the $\rho_A$  \eqref{eq:rhoCylinderI} is a projector up to a normalization factor.
We have\begin{equation}\label{eq:rhorhoCylinderI}
\begin{aligned}
\rho_A\rho_A=
&\sum_{\substack{g_v,h,h'\in G,\\ k_1,k_3\in K_1,\\ k_2,k_4\in K_2}}\ \sum_{\substack{ \bar g_v,\bar h,\bar h'\in G,\\ \bar k_1,\bar k_3\in K_1,\\ \bar k_2,\bar k_4\in K_2}}
\ketbra{k_uh,g_vh}{\bar k_uh',\bar g_vh'}\braket{k_u\bar h,g_v\bar h}{\bar k_u\bar h',\bar g_v\bar h'}\\
=&\beta\sum_{\substack{g_v,h,h'\in G,\\ k_1,k_3\in K_1,\\ k_2,k_4\in K_2}}
\ketbra{k_uh,g_vh}{k_uh',g_vh'}\\
=&\beta \rho_A.
\end{aligned}
\end{equation}
Here, the constant factor $\beta$ is
\begin{equation}\label{eq:rhorhoCylinderIBeta}
\begin{aligned}
\beta=
&\quad
\sum_{\substack{\bar g_v,\bar k_1,\bar k_3,\\ \bar k_2,\bar k_4,\bar h,\bar h'}}
\sum_{c\in K}\delta_{c,\bar{h} {\bar{h'}}^{-1}}
\delta_{c,\bar g_v g_v^{-1}}\delta_{c,\bar k_u k_u^{-1}}\\
=& \sum_{\bar h \in G}\sum_{c\in K} 1\\
=&|G| |K|,
\end{aligned}
\end{equation}
where $c$ runs over $K=K_1\cap K_2$. To see the second equality above, we note that for example, 
\begin{align*}
\sum_{c\in G} \sum_{\bar k_u}\delta_{c,\bar k_u k_u^{-1}} &=\sum_{c\in G}\ \sum_{\bar k_1, \bar k_2 \in K_1, \bar k_3, \bar k_4 \in K_2}\delta_{c,\bar k_1 k_1^{-1}} \delta_{c,\bar k_2 k_2^{-1}} \delta_{c,\bar k_3 k_3^{-1}} \delta_{c,\bar k_4 k_4^{-1}}\\
&=\sum_{c\in K}\  \sum_{\bar k_1, \bar k_2 \in K_1, \bar k_3, \bar k_4 \in K_2}\delta_{c,\bar k_1 k_1^{-1}}\delta_{\bar k_1 k_1^{-1}, \bar k_3 k_3^{-1}} \delta_{\bar k_2 k_2^{-1}, \bar k_3 k_3^{-1}} \delta_{\bar k_2 k_2^{-1}, \bar k_4 k_4^{-1}} \\
&=\sum_{c\in K}\ \sum_{\bar k_1, \bar k_2 \in K_1, \bar k_3, \bar k_4 \in K_2}\delta_{\bar k_1,c k_1} \delta_{\bar k_2,ck_2} \delta_{\bar k_3, ck_3} \delta_{\bar k_4, ck_4} \\
&= \sum_{c\in K}1,
\end{align*}
which also constrains $c\in K$. We can evaluate the trace of the projection operator $\beta^{-1}\rho_A$:
\begin{equation}\label{eq:tracerhoCylinderI}
W:=\mathrm{tr}(\beta^{-1}\rho_A)
=\frac{|G|^L|K_1|^2|K_2|^2}{|K|},
\end{equation}
where $L$ is the total number of bulk vertices (those not on PB) of the EB.
Hence the EE:
\begin{equation}\label{eq:EECylinderIappendix}
S_E=\log W=\log \frac{|G|^L|K_1|^2|K_2|^2}{|K|}.
\end{equation}

\section{Reduced density matrix in cylinder case II}
\label{sec:rhoCylinderII}
We show that the $\rho_A$  \eqref{eq:rhoCylinderI} is a projector up to a normalization factor when the gauge group $G$ is Abelian.
We have\begin{equation}\label{eq:rhorhoCylinderII}
\begin{aligned}
\rho_A\rho_A=
&\sum_{k_1k_2g_v,g'_v,h,h'}\ \sum_{\bar k_1\bar k_2\bar g_v,\bar g'_v,\bar h,\bar h'}\ketbra{g_vh,g'_vh}{\bar g_v\bar k_1\bar h',\bar g'_v\bar k_2\bar h'}
\braket{g_vk_1h',g'_vk_2h'}{\bar g_v\bar h,\bar g'_v\bar h}
\\
=
&\sum_{\substack{k_1 k_2 g_v,\\ g'_v,h,h'}}\ \sum_{\substack{ \bar k_1\bar k_2\\ \bar g_v,\bar g'_v,\bar h,\bar h'}}
\sum_{c\in G}
\ketbra{g_vh,g'_vh}{\bar g_v\bar k_1\bar h',\bar g'_v\bar k_2\bar h'}
\delta_{\bar h{h'}^{-1},c}\delta_{{\bar g_v}^{-1}g_v,ck_1^{-1}}\delta_{{{\bar g}_v}^{\prime-1}g'_v,ck_2^{-1}}
\\
=
&\sum_{g_v,g'_v,h,k_1k_2,h',\bar k_1\bar k_2,\bar h'}
\ketbra{g_vh,g'_vh}{g_vk_1h'\bar k_1\bar h',g'_vk_2h'\bar k_2\bar h'}.
\end{aligned}
\end{equation}
When $G$ is Abelian, we can define a projector
\begin{equation}\label{eq:CylinderIIAbelianrhoA}
\bar\rho_A=\beta^{-1} \rho_A,
\end{equation}
with the constant factor
\begin{equation}\label{eq:rhorhoCylinderIIBeta}
\begin{aligned}
\beta=
\sum_{\bar k_1, \bar k_2, \bar h'}1
=|G|^2|K_1||K_2|.
\end{aligned}
\end{equation}
We can evaluate the trace of the projector $\beta^{-1}\rho_A$:
\begin{equation}\label{eq:tracerhoCylinderII}
W=\mathrm{tr}( \bar \rho_A)
=\frac{1}{|G|^2|K_1||K_2|}
\sum_{k_1, k_2, g_v, g'_v, h, h'}\delta_{h,k_1h'}\delta_{h,k_2h'}
=\frac{|G|^{L-1}|K|}{|K_1||K_2|},
\end{equation}
where $L$ is the total number of vertices on the EB.
The EE  in the Abelian case then is
\begin{equation}\label{eq:EECylinderIIappendix}
S_E=\log W=\log \frac{|G|^{L-1}|K|}{|K_1||K_2|}.
\end{equation}

\section{Some useful identities}

Let $G$ be a finite group and $K$ a subgroup of $G$. Denote by $\{j\}$ all unitary irreducible representations of $G$, and by $\{p\}$ those of $K$ respectively. Denote the representation matrices by $\rho^j(g)$ and $R^p(k)$ for $g\in G$ and $k\in K$ respectively.

Obviously any representation $j$ of $G$ is automatically a (not necessarily irreducible) representation of $K$ and can be decomposed into a direct sum of certain irreducible representations of $K$, namely,
\begin{equation}\label{eq:Njp}
j=\oplus_p N^j_p p.
\end{equation}
More precisely, the decomposition means that there exists a unitary transformation $U^j$ rendering $U^j\rho^j(k){U^j}^{\dagger}$ a direct sum of $R^p(k)$ for certain $p$'s. The multiplicities $N^j_p$ count the numbers $R^p(k)$ appearing in the direct sum. We write $U^j\rho^j(k){U^j}^{\dagger}=\oplus_p N^j_p R^p(k)$. It may be useful to write down $U^j$ more explicitly: 
\begin{equation}\label{eq:UU}
\rho^j(k)=\sum_{p}\sum_{\alpha\in N^j_p}{U^j_{p,\alpha}}^\dagger R^p(k)U^j_{p,\alpha}
\end{equation}
and
\begin{equation}\label{eq:UUid}
U^j_{p,\alpha}{U^j_{p',\alpha'}}^\dagger=\delta_{j,j'}\delta_{\alpha,\alpha'}\mathrm{id}_p.
\end{equation}
Equation \eqref{eq:UU} is presented graphically as
\begin{equation}\label{eq:UUpic}
\Decomposej
\end{equation}
The multiplicities $N^j_p$ compute as
\begin{equation}\label{eq:multiplicity}
N^j_p=\frac{1}{|K|}\sum_{k\in K}\mathrm{tr}\rho^j(k)\overline{\mathrm{tr} R^p(k)}.
\end{equation}
Equation \eqref{eq:UU} leads to a useful identity:
\begin{equation}\label{eq:rhokrhok}
\frac{1}{|K|}\sum_{k\in K}\rho^j_{mn}(k)\overline{\rho^{j'}_{m'n'}(k)}=
\sum_{p}\frac{1}{d_p}\sum_{\alpha\beta}({U^j_{p\alpha}}^\dagger U^{j'}_{p\beta})_{m'm}({U^{j'}_{p\beta}}^\dagger U^{j}_{p\alpha})_{nn'}.
\end{equation}

\section{Examples}\label{sec:example}

In Section \ref{subsubsec:CylIIfusion}, we computed as an example the EE in cylinder case II when $G=S_3$. Here, we explain some basic setup for $S_3$ and its representations.

Denote the group elements of $S_3$ by $g=1,\dots,6$, with the generators $2,4$ satisfying $2\cdot 2\cdot 2 =1$, $4\cdot 4=1$, and $4\cdot 2\cdot 4=2\cdot 2$. Denote the rest by $3=2\cdot2$, $5=2\cdot 4$, and $6=2\cdot 5$.

The group $S_3$ has three irreducible representations. A representative set of the irreducible representations is
\begin{equation}\label{eq:reps}
\begin{array}{cccc}
& g=1 & g=2 & g=4 \\
\rho ^1(g) & \left(
\begin{array}{c}
1 \\
\end{array}
\right) & \left(
\begin{array}{c}
1 \\
\end{array}
\right) & \left(
\begin{array}{c}
1 \\
\end{array}
\right) \\
\rho ^2(g) & \left(
\begin{array}{c}
1 \\
\end{array}
\right) & \left(
\begin{array}{c}
1 \\
\end{array}
\right) & \left(
\begin{array}{c}
-1 \\
\end{array}
\right) \\
\rho ^3(g) & \left(
\begin{array}{cc}
1 & 0 \\
0 & 1 \\
\end{array}
\right) & \left(
\begin{array}{cc}
-\frac{1}{2} & -\frac{\sqrt{3}}{2} \\
\frac{\sqrt{3}}{2} & -\frac{1}{2} \\
\end{array}
\right) & \left(
\begin{array}{cc}
\frac{1}{2} & \frac{\sqrt{3}}{2} \\
\frac{\sqrt{3}}{2} & -\frac{1}{2} \\
\end{array}
\right) \\
\end{array}.
\end{equation}
The representation matrices for other group elements can be obtained by the group multiplications. In this setting, Eq. \eqref{eq:Netaexample} was obtained by direct computation.

%%%%%%%%%%%%%%%%%%%
%%%%%%%%%%%%%%%%%%%
%%%%%%%%%%%%%%%%%%%

\acknowledgments
We are grateful to Ling-Yan Hung for helpful discussions and critical comments on the manuscript. YTH thanks Yong-Shi Wu for inspirations and appreciates Wenqing Zhang for his support. YDW thanks IQC for hospitality during his visit, where this paper is finalized, and the Perimeter Institute for hospitality during his visit, where part of the work was done; he is also supported by the Shanghai Pujiang Program No. 17PJ1400700 and the NSF grant No. 11875109.

\bibliographystyle{apsrev4-1}
\bibliography{StringNet}
\end{document}